\documentclass{article}[12pt,a4paper]
\pdfoutput=1
\usepackage{graphicx}  
\usepackage{dcolumn}   
\usepackage{bm}        
\usepackage{amssymb}   
\usepackage[]{amsmath,amssymb}
\usepackage{graphics,epsfig}
\usepackage[latin1]{inputenc}
\usepackage[T1]{fontenc}
\usepackage{amsfonts}
\usepackage{latexsym}
\usepackage[english]{babel}
\usepackage[usenames, dvipsnames]{color}
\usepackage{booktabs}
\usepackage{multirow}
\newcommand{\be}{\begin{equation}}
\newcommand{\ee}{\end{equation}}
\newcommand{\beq}{\begin{equation}}
\newcommand{\eeq}{\end{equation}}
\newcommand{\bea}{\begin{eqnarray}}
\newcommand{\eea}{\end{eqnarray}}
\newcommand{\nn}{\nonumber}
\def\be{\begin{equation}}
\def\ee{\end{equation}}
\def\ba{\begin{eqnarray}}
\def\ea{\end{eqnarray}}

\begin{document}

\title{Intermediate scalings for Solv, Nil and $SL_2({ \cal R})$ black branes}

\author{Ra\'{u}l E. Arias$^{a}$\footnote{rarias@sissa.it}, Ignacio Salazar Landea$^{b, c}$\footnote{peznacho@gmail.com}}
\maketitle

\begin{center}

{\sl $^a$SISSA\\ 
Via Bonomea 265, 34136 Trieste, Italy} 

~

{\sl $^b$ Centro At\'omico Bariloche,\\
8400-S.C. de Bariloche, R\'{\i}o Negro, Argentina}

~

{\sl $^c$ Instituto de F\'{\i}sica de La Plata - CONICET\\
C.C. 67, 1900 La Plata, Argentina.}

\end{center}

~

~

\begin{abstract}

In this work we will study black brane solutions that are not translationally invariant in the spatial directions along which it extends. Instead, we require homogeneity, which still allows points along the spatial directions to be related to each other by symmetries. We find Einstein-Maxwell-AdS black hole solutions whose near horizon geometry correspond to Solv (Bianchi $V1_{-1}$), Nil (Bianchi $II$) or $SL_2({\cal R})$ (Bianchi $VIII$). Interestingly we observe that at intermediate temperatures our solutions have an scaling regime where different space time directions scale differently. We also compute the DC conductivities for these charged solutions and study how they scale in this intermediate regime.

\end{abstract}

\tableofcontents


\section{Introduction}

Quantum critical points and novel phases of matter with unconventional scalings are systems whose degrees of freedom are believed to be in a strongly coupled regime. Then, holography is a good playground with many tools to study the dynamics of such systems. Moreover there is an effort to classify phase structures and in particular in the deep infrared (IR) regime. The result of such efforts was the appearance of novel RG flows geometries with intermediate scalings which are useful to model the behaviour of conformal field theories around quantum critical points. In this work we will show new intermediate regimes in a class of Einstein-Maxwell theories that break translational invariance in a particular way that still preserves homogeneity.

In the context of the AdS/CFT approach to study condensed matter physics there has been significant recent interest in the construction of black hole solutions dual to CFTs deformed by operators that break translational invariance. This is because these systems allow momentum to dissipate giving place to study more realistic transport properties. The breaking of translational invariance in holography is a complicated issue because typically this drive us to solve partial differential equations  \cite{Horowitz:2012ky,Horowitz:2012gs,Horowitz:2013jaa}.

One way\footnote{Another powerfull way to incorporate the effect of momentum dissipation and still work with ODEs is to effectively add the effect of inhomogeneities by giving a mass to the graviton  \cite{Vegh:2013sk,Blake:2013bqa,Blake:2013owa,Amoretti:2014zha,Amoretti:2014mma}} to bypass this complication for theories in $d=4$ dimensions is generalizing the generators of the translational symmetry to a Bianchi symmetry, where they do not commute and then explointing this new symmetry to get ODEs \cite{Donos:2012js,Iizuka:2012iv,Cadeau:2000tj,Hassaine:2015ifa,Arias:2017yqj,Donos:2014gya}. The Thurston geometrization conjecture gives a classification of geometries that are allowed to be the near horizon limit of a black hole solution. There are just 8 of such geometries (Euclidean space $E_3$, the three sphere $S^3$, the hyperbolic space $H_3$, the products $S_1\times H_2$ and $S_1\times S_2$, Nil geometry, Solv geometry and lastly the universal cover $SL_2(\cal{R})$) and the remaining ones must be isometric to one of them. In the present work we are going to elaborate on the direction of \cite{Arias:2017yqj}  and we will study black holes where the horizon have Bianchi $VI_{-1}$ (Solv), Bianchi $II$ (Nil) or Bianchi $VIII$ ($SL_2({ \cal R})$) symmetries
\bea
Solv: \ \ \ \ \ \ d\bar{s}^2&=& e^{2kz}dx^2+ e^{-2kz}dy^2+dz^2 \nn \\
Nil: \ \ \ \ \ \ d\bar{s}^2&=& dx^2+ dy^2+(dz-kxdy)^2 \,,  \nn \\
SL_2({ \cal R}): \ \ \ \ \ \ d\bar{s}^2&=& \frac1{k^2x^2}( dx^2+ dy^2 )+ \left(dz+\frac{dy}{kx}  \right)^2\,,
 \label{bianchist}
\eea
that we will generically write as 
\be
d\bar{s}^2=\sum_{i=1}^3 \omega_i^2\,.
\ee

For the solv geometry $\omega_1=e^{kz}dx$, $\omega_2=e^{-kz}dy$, $\omega_3=dz$, for the Nil solutions $\omega_1=dx$, $\omega_2=dy$, $\omega_3=dz-kxdy$ while for the $SL_2({ \cal R})$ we have $\omega_1=dx/kx$, $\omega_2=dy/kx$, $\omega_3=dz+dy/kx$.

In this setup we will obtain solutions of the Einstein-Maxwell-AdS action in five dimensions
\be 
\tilde S=\int d^5 x \sqrt{-g}\left(  R -12- \frac14 F_{\mu \nu}F^{\mu\nu}\right)+\tilde S_{bdy}\,,
 \label{EM}
\ee
where $\tilde S_{bdy}$ corresponds to some boundary action, solving the following equations of motion
\bea
D_\mu F^{\mu\nu}&=&0\,,\nn \\ 
R_{\mu \nu}- \frac12 R\ g_{\mu\nu}-6 \  g_{\mu\nu} & =& -\frac14 F_{\alpha \beta}F^{\alpha\beta} g_{\mu\nu} +  F_{\mu \alpha}F^\alpha_{\,\,\, \nu}\,.
\label{EMeoms}
\eea
This action corresponds to the universal sector of many holographic models. In this context we will build charged black holes solutions that will be dual to a finite temperature CFT  at a finite chemical potential living in the spacetimes defined in (\ref{bianchist}). 

This paper will elaborate on the solutions found in \cite{Arias:2017yqj} by relaxing the ansatz for the metric. This will allow us to find solutions that are isotropic in the UV giving a more natural framework for an holographic interpretation. We can now think our solutions as the dual of a four dimensional CFT deformed by placing the theory on a Bianchi manifold.

As recently reported in \cite{Horowitz:2012gs,Donos:2017ljs,Donos:2017sba}, breaking translational symmetry gives rise to new scalings at the IR or at intermediate scales. We will study these new scalings by constructing a family of black hole solutions by varying how badly is the translational symmetry broken. We will see that for neutral black holes new scalings will appear near the horizon of low reduced temperature $T/k$ black holes. In these new scalings, different spatial coordinates will typically scale differently. 
When considering charged solutions, these new scalings become intermediate scalings and our near horizon solutions look like $AdS_2 \times Thurston$.

Recently it was argued \cite{Hartnoll:2015sea} that many properties of strange metals could be explained from a simple scaling theory. Then it is interesting to understand the mechanisms that make a theory to develop non conventional scalings. In this context, we show evidence that breaking the translational invariance on the universal sector of any theory with a gravity dual forces the theory to develop new scalings at low or intermediate temperatures.

Having the translational symmetry broken, the dual field theories will have finite DC transport coefficients. We  can read such coefficients directly from the horizon metric following the method developed in \cite{Donos:2014cya}. We see that the DC conductivities also reflect the intermediate scalings at intermediate temperatures \cite{Bhattacharya:2014dea}.

This paper organizes as follows. On each of the following sections we show solutions to the Einstein-Maxwell equations with different Thurston geometry horizons. Then we proceed to compute the thermoelectric transport coefficients in the direction in which the translational symmetry is broken. In Section \ref{sec2} we deal with Solv horizons, in Section \ref{sec3} with Nil, while in Section \ref{sec4} with $SL_2({ \cal R})$ horizons. Finally in Section \ref{sec5} we summarize our results and discuss possible future directions.

\section{Solv black holes with intermediate scaling}
\label{sec2}

In this section we will work with black holes with solvgeometry horizons.

\subsection{Solutions}
\label{sec2a}

We will consider the following ansatz for the gauge and metric fields
\bea
A&=&A_t(r)\  dt\, ,   \nn\\
ds^2&=&- r^2 f^2(r)g(r) dt^2 + \frac1{r^2 g(r)} dr^2 +r^2 h^2(r)\left( e^{2 k z }dx^2 + e^{-2kz} dy^2\right)+ \frac{r^2}{ h^4(r)} dz^ 2\,.
\eea

Explicitly, the equations of motion \eqref{EMeoms} read
\bea
g'-\frac4r + \frac{2 k^2 h^4}{3r^3} + \frac{A_t'^2}{3rf^2}  + \left(\frac4r + \frac{2r h'^2}{h^2}  \right)g=0\,, \nn \\
3 r^4 f h'' + \left(12+3 g-\frac{A_t'^2 }{ g^2 } \right)r^3 h'-2 k^2 \left(h^5+ r h^4 h'\right)- \frac{3r^4\,g\,h'^2}{h}=0\,, \nn\\
h^2f'-2r f h'^2=0\,, \nn\\
A_t''+\left( \frac3r- \frac{f'}{f}   \right)A_t'=0\,,
\eea
with primes denoting radial derivative $'=\partial_r$ and where we drop the $r$ dependence of the functions metric and gauge functions.

We will look for black hole solutions to these equations by integrating out the fields from the near horizon
\bea
A_t(r)&\simeq & a_{t_1}(r-r_h)-\frac{a_{t_1} \left(12 f_0^2 h_0^4 k^2 r_h^2 \left(a_{t_1}^2-12 f_0^2\right)+3 r_h^4 \left(a_{t_1}^2-12
   f_0^2\right)^2+4 f_0^4 h_0^8 k^4\right)}{2 r_h \left(r_h^2 \left(a_{t_1}^2-12 f_0^2\right)+2 f_0^2 h_0^4
   k^2\right)^2}(r-r_h)^2+\ldots, \nn\\
f(r)&\simeq & f_0+\frac{8 f_0^5 h_0^8 k^4}{r_h \left(r_h^2 \left(a_{t_1}^2-12 f_0^2\right)+2 f_0^2 h_0^4 k^2\right)^2}(r-r_h)+\ldots,\nn\\
g(r)&\simeq &\frac{r_h^2 \left(12-\frac{a_{t_1}^2}{f_0^2}\right)-2 h_0^4 k^2}{3 r_h^3}(r-r_h)+\ldots,\nn\\
h(r)&\simeq & h_0-\frac{2 f_0^2 h_0^5 k^2}{r_h^3 \left(a_{t_1}^2-12 f_0^2\right)+2 f_0^2 h_0^4 k^2 r_h}(r-r_h)+\ldots, \label{NHsolv}
\eea
towards the boundary
\bea
g(r)&\approx& 1 - \frac{h_{\infty}^4k^2}{3r^2}+\frac{g^{\infty}_4}{r^4}+\frac{2h_\infty^8k^4\log r}{r^4}+\dots\,, \nn\\
h(r)&\approx& h_\infty-\frac{h_\infty^5k^2}{6r^2}+\frac{h_4^\infty}{r^4}+\frac{h_\infty^9k^4\log r}{9r^4}+\dots\,,\nn\\
f(r)&\approx& f_\infty-\frac{f_\infty h_\infty^8k^4}{r^4}+\dots\,,\nn\\
A_t(r)&\approx&  \mu+\frac{\rho}{r^2}-\frac{2\rho f_\infty h_\infty^8 k^4}{108f_\infty r^6}+\dots\,.
\eea
The parameters $a_{t_1}$, $f_0$, $h_0$ are the independent coefficients of the functions expanded around the horizon and $h_\infty$, $h_4^\infty$, $g_\infty$, $g_4^\infty$,$f_\infty$ are the corresponding ones when the expansion is around the UV.
This boundary conditions imply that our black hole solutions have the same scaling for the metric (towards the boundary) in all the directions which is an important difference with respect to previous works \cite{Cadeau:2000tj,Hassaine:2015ifa,Arias:2017yqj}. The $\log$ terms are associated with an anomalous scaling of physical quantities due to
the conformal anomaly. As expected, they go to zero when considering $k\rightarrow0$.

The thermodynamics of the black hole are given by its temperature $T$ and entropy density $s$
\bea
T&=&\frac{1}{4\pi}\left(\frac{g'_{tt}}{\sqrt{g_{rr}g_{tt}}}\right)_{r=r_h}=\frac{\left(\frac{a_{t_1}^2 }{3 f_0}-4f_0\right)r_h+\frac{2 f_0 h_0^4 k^2}{3 r_h}}{4 \pi }, \nn \\ 
s&=&2\pi A_h=2\pi r_h^3,
\eea
with $A_h$ denoting the area of the black hole horizon. We find a family of solutions that we characterize by the dimensionless parameters $T/k$ and $\mu/k$.

\begin{figure}[!htb]
\begin{center}  
\includegraphics[scale=0.63]{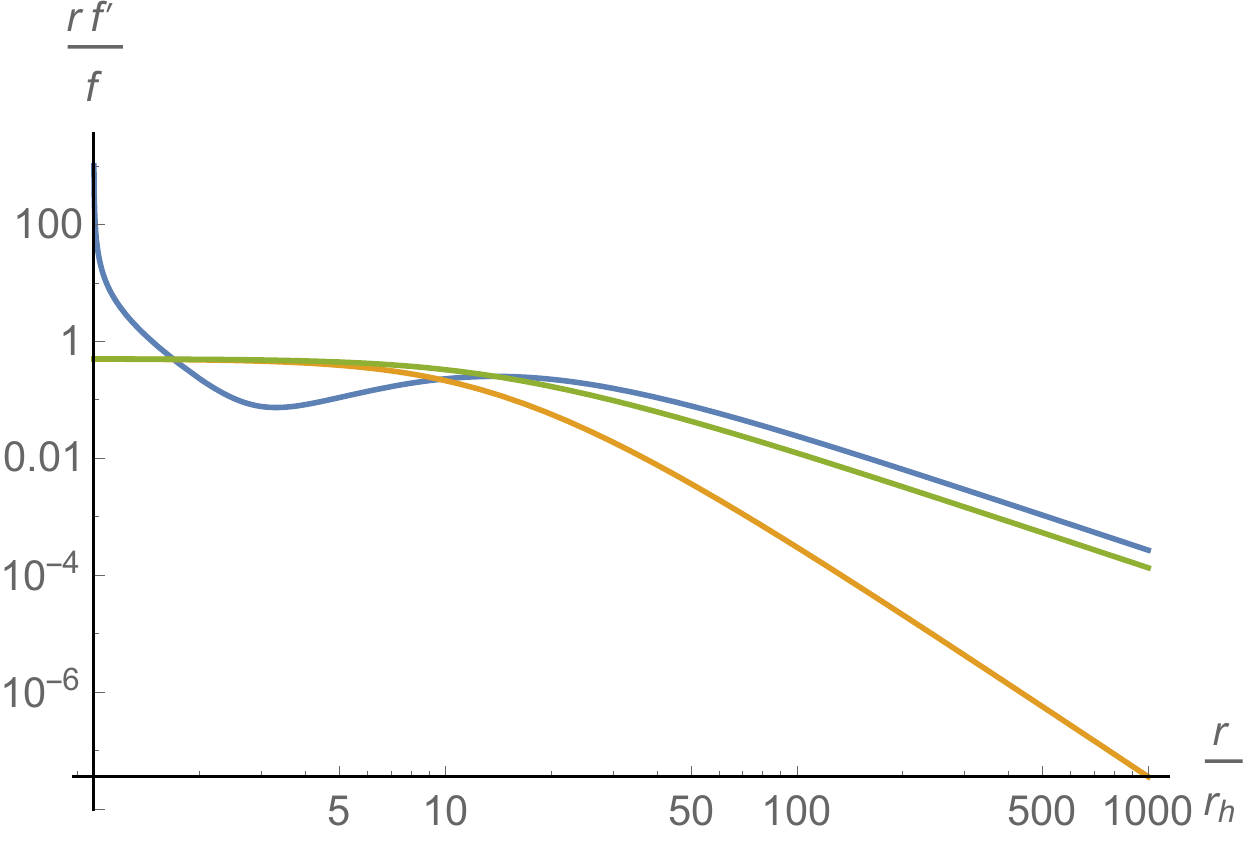}\hfill\includegraphics[scale=0.63]{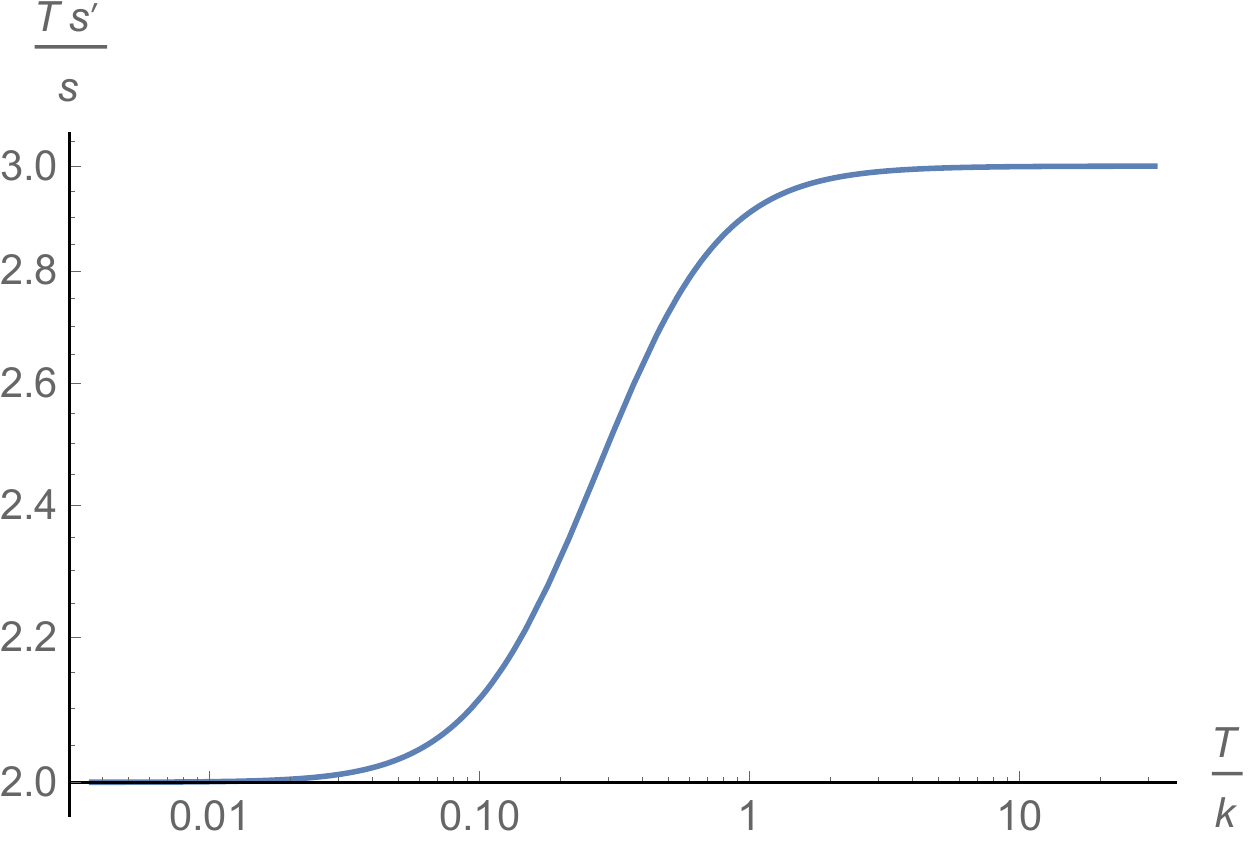}
\caption{Left: Typical profile for $f(r)$ (blue), $g(r)$ (orange) and $h(r)$ (green). The combination $rf'/f$, where $f$ corresponds to any of the aforementioned functions, was chosen to make manifest the emergence of new scalings towards the IR. The plot corresponds to a solution with $T/k=0.00367$ and $\mu/k=0$. Right: $T s'/s$ as a function of the reduced temperature $T/k$.  \label{fig1}}
\end{center}  
\end{figure}

For simplicity, let us begin by studying solutions with $\mu=0$, corresponding to neutral black holes.
Integrating the equations of motion we find a family of solutions with different $T/k$. In the right panel of 
Figure \ref{fig1} we show how the entropy of the corresponding solutions scales with respect to the Temperature. As we can see, for high enough temperatures we see the expected $CFT_4$ related scaling $s/k^3\sim (T/k)^3$. On the other hand,
for low enough temperatures we find a new scaling $s/k^3\sim (T/k)^2$. To understand better the nature of this new scaling, we must study in detail the behavior of the metric fields.

In the left panel of Figure \ref{fig1} we show a typical plot for the metric fields as a function of the radial coordinate, when considering a solution with  low enough $T/k$.
From the numerics we extract that the fields $g$ and $h$ of the metric get a new scaling at small $r/r_h$, such that $g\sim h \sim r^{1/2}$. Plugging this again into the metric one can extract the following scalings 
\bea
t \rightarrow \lambda\, t\,,\,\,\,\, x\rightarrow \lambda\, x\, , \,\,\,\, y\rightarrow \lambda\, y \,, \,\,\,\, z\rightarrow z \,.
\eea

This means that we have anisotropic scalings but these are not of the standard Lifshitz type, since is one of the spatial directions the one scaling differently than the others.

It seems tempting to associate the intermediate scaling geometry with
\bea
A&=&0  \, ,   \nn\\
\frac43 ds^2&=&- r^2 dt^2 + \frac1{r^2 } dr^2 +r^2 \left( e^{2 kz }dx^2 + e^{-2kz} dy^2\right)+ \frac{2k^2}{3 } dz^ 2\,.\label{solvinter}
\eea
This is an exact solution to the Einstein-Maxwell equations of motion.
These exact solutions were previously studied in \cite{Cadeau:2000tj,Hassaine:2015ifa,Arias:2017yqj}. Here we show how this solutions emerge naturally in the IR regime when studying black hole solutions with solvgeometry horizons that are asymptotically isotropic.

\begin{figure}[!htb]
\begin{center}  
\includegraphics[scale=0.63]{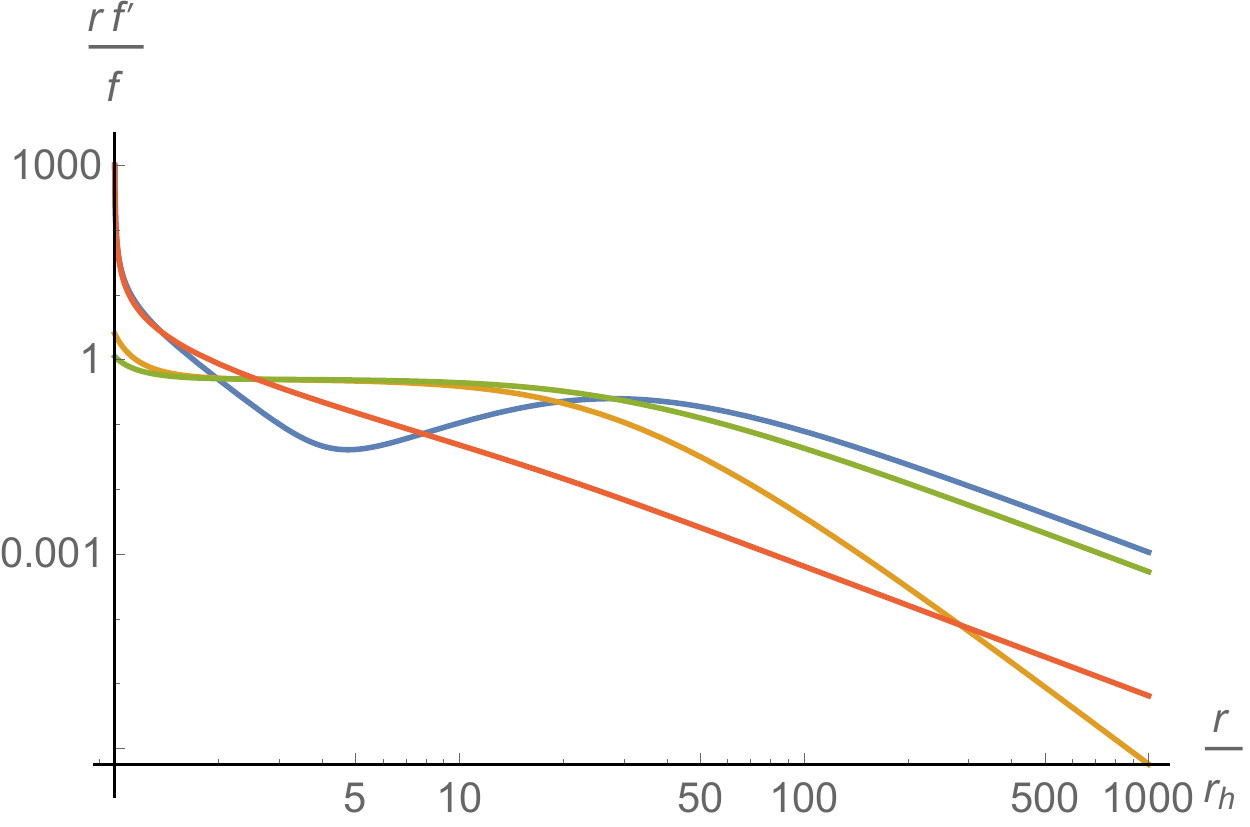}\hfill\includegraphics[scale=0.63]{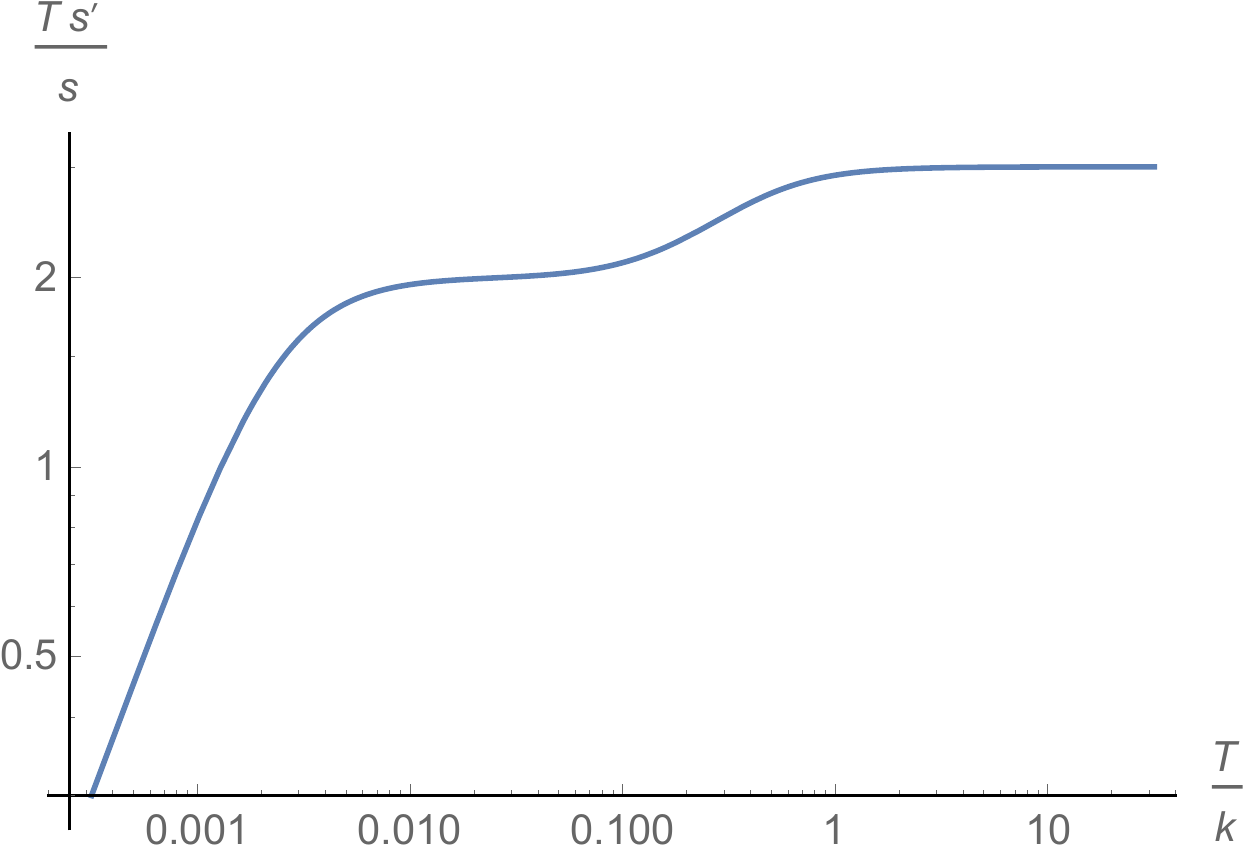}
\caption{Left: Typical profile for $f(r)$ (blue), $g(r)$ (orange), $h(r)$ (green) and $A_t(r)$ (red). The combination $rf'/f$, where $f$ corresponds to any of the aforementioned functions, was chosen to make manifest the emergence of new scalings towards the IR. The plot corresponds to a solution with $T/k=0.0003185$ and $\mu/k=0.01$. Right: $T s'/s$ as a function of the reduced temperature $T/k$.  \label{fig2}}
\end{center}  
\end{figure}

If we now turn to charged black hole solutions, the IR scaling becomes an intermediate scaling, and the geometry flows into $AdS_2\times Solv$ in the deep near horizon regime, as expected for a charged black hole. In Figure \ref{fig2} we show typical profiles for the fields and the scaling of the entropy for a family of solutions with fixed $\mu/k=0.01$.

It is important to notice that the value of the chemical potential in units of $k$ must be small enough to have intermediate scaling solutions.  If it is not, then the $AdS_{2}\times solv$  geometry will appear at larger values of the radius and will ruin the intermediate scaling. In other words, we need $\mu/k$ small enough so that the intermediate scaling has enough room to appear. The same argument will hold for the Nil and $SL_2({ \cal R})$ cases of the next sections.

\subsection{Finite conductivities from Solv horizons}
\label{sec2b}

Consider a system at equilibrium at finite chemical potential and temperature.
Adding a small electric field $E_i$ or thermal gradient $\nabla_i T$ will induce an electric current $J ^i$ and a heat current $Q^i=T^{ti}-\mu J^i$, where $T^{ij}$ is the stress tensor of the dual field theory. At linearized order, the response is controlled by the Ohm/Fourier law 
\bea
\left(\begin{matrix}
 J \\
 Q
\end{matrix}\right)=\left(\begin{matrix}
   \sigma    & \alpha T \\
  \bar\alpha T    &\bar\kappa T
\end{matrix}\right) \left(\begin{matrix}
   E \\
- \nabla T/T
\end{matrix}\right)\,,
\label{ohm}
\eea
defining the electric conductivity $\sigma$, the thermoelectric conductivities $\alpha$, $\bar\alpha$ and the thermal conductivity $\bar\kappa$.

Systems with translation invariance and finite charge density have an infinite $DC$ conductivity. Nonetheless, in the directions where the translation invariance is broken, we expect a finite $DC$  conductivity. That will be the $z$ direction in our solvgeometry charged black holes or the $x$ direction in our nilgeometry and $SL_2({\cal R})$ charged black holes. We will read then the coefficients of the matrix (\ref{ohm})  from horizon data, following the method developed in \cite{Donos:2014cya}.

The holographic dictionary gives us the expressions for the electric and heat current in the dual field theory \cite{Liu:2017kml,Donos:2017oym}
\bea
J&=&\sqrt{-g} F^{ri}\,,\nn\\ 
Q&=&\sqrt{-g}G^{ri} + J A_t,\label{JQ}
\eea
where the tensor $G^{\mu\nu}$ reads
\be 
G^{\mu\nu}=\nabla^\mu k^\nu +\frac{1}{3}k^{[\mu}F^{\nu]\sigma}A_\sigma, 
\ee
$k=\partial_t$ and the index $i$ denotes the direction on which the electric field is applied.

\subsubsection{Computing $\sigma$ and $\bar\alpha$}

Let us begin with the computation of the electric conductivity, and one of the thermoelectric conductivities $\bar\alpha$. To do that we will study linear response under the following fluctuations of the metric and gauge field 

\bea
\delta A&=&(-E t+\delta a_z(r))dz\nn\\
\delta ds^2&=&h_{tz}(r)dtdz+h_{rz}(r)drdz,
\eea
where the constant $E$ is the applied (DC) electric field. From the Maxwell equations we obtain the following expression for the electric current $J$

\be
J=\frac{r h^4 \left(h_{tz}A_t'+2 f^2 r^2 g\, a_z'\right)}{2 f}\,.
\ee
In order to have clean expressions we will not write the $r$ dependence of the functions. It is easy to check that the equation of motion for $\delta a_z$ is equivalent to $\partial_r J=0$. This allows us to evaluate the current at any value of the radial coordinate, in particular we want to express the transport coefficients as function of IR data, i.e, the position of the horizon of the black hole. 

Using the Einstein equations  we obtain that the heat current can be written as
\be
 Q=\frac{r h^4 \left(2 A_t\, h_{tz}\,A_t'+r f^2 \left(g \left(4 r A_t a_z'+r h_{tz}'-2
   h_{tz}\right)-r h_{tz} g'\right)-2 r^2 f g \,h_{tz} f'\right)}{4 f}\,.
\ee 
Again, we can see that $\partial_r Q=0$ and then we can evaluate it on $r=r_h$. The remaining Einstein equation is

\bea
h_{rz}&=&-\frac{r^2 E  A_t'}{k^2 f^2 g h^4}\,.\label{eomhrz1}
\eea

For a free falling observer the horizon of a black hole is a regular place, then the electromagnetic field must be regular there. Using Eddington-Finkelstein coordinates $dv=dt+\sqrt{\frac{g_{rr}}{g_{tt}}}\,dr$ and asking for regularity of the fluctuations at $r=r_h$ we can obtain the near horizon behaviour of $h_{tz}$ and $\delta a_z$. Using the expansions \eqref{NHsolv} and the near horizon limit of \eqref{eomhrz1} we can express the electric current and the Heat kernel as function of IR data as

\bea 
J&=&\frac{1}{2} r_h^3 E  \left(-2 h_0^4+\frac{a_{t_1}^2}{f_0^2 k^2}\right)\,,\nn\\
Q&=&-\frac{a_{t_1} r_h^2 E  \left(r_h^2 \left(12-\frac{a_{t_1}^2}{f_0^2}\right)-2 h_0^4 k^2\right)}{12 k^2}\,.
\eea

From this we can compute the conductivities $\sigma$ and $\bar\alpha$
\bea
\sigma=\frac{\partial J}{\partial E}&=&\frac{1}{2} r_h^3 \left(2 h_0^4+\frac{a_{t_1}^2}{f_0^2 k^2}\right)\,,\nn\\
\bar\alpha=\frac{1}{T}\frac{\partial Q}{\partial E}&=&\frac{\pi  a_{t_1} r_h^3}{f_0 k^2}\,.
\eea

\begin{figure}[!htb]
\begin{center}  
\includegraphics[scale=0.63]{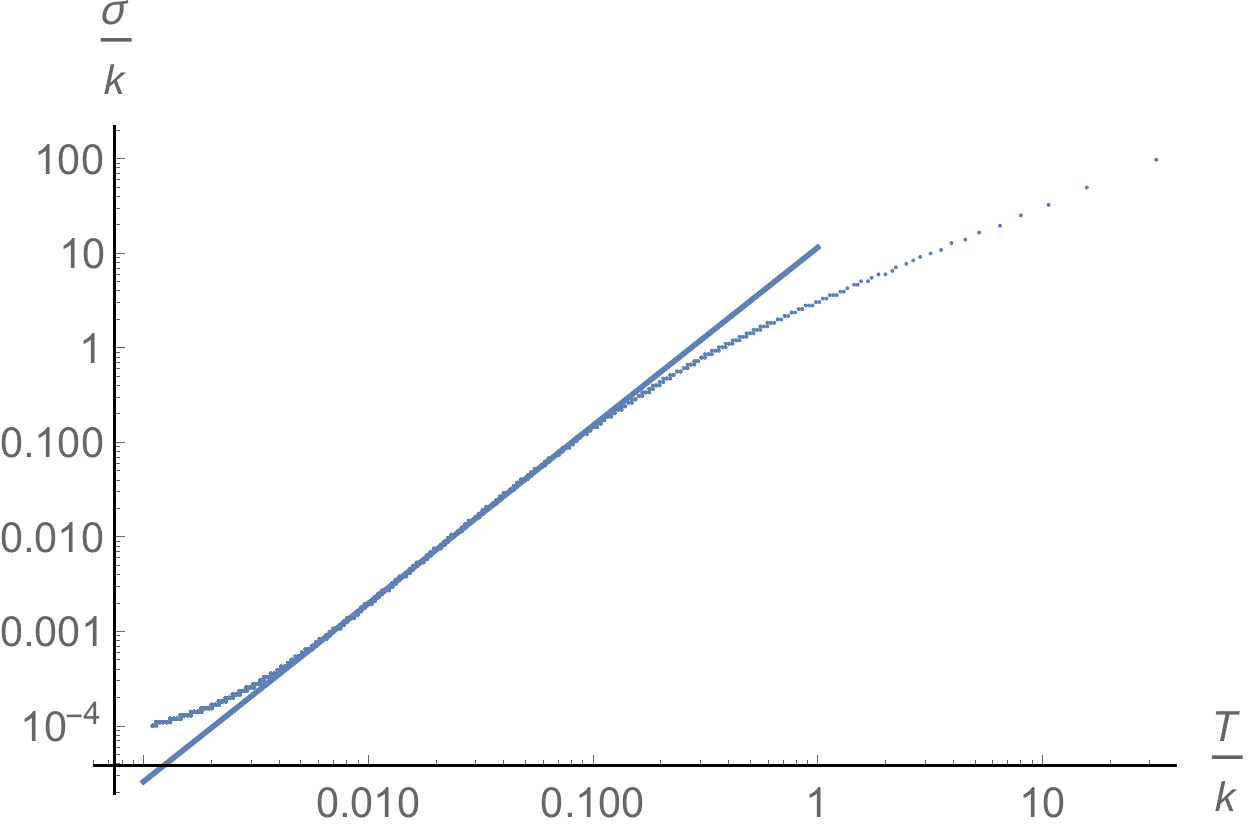}\hfill\includegraphics[scale=0.63]{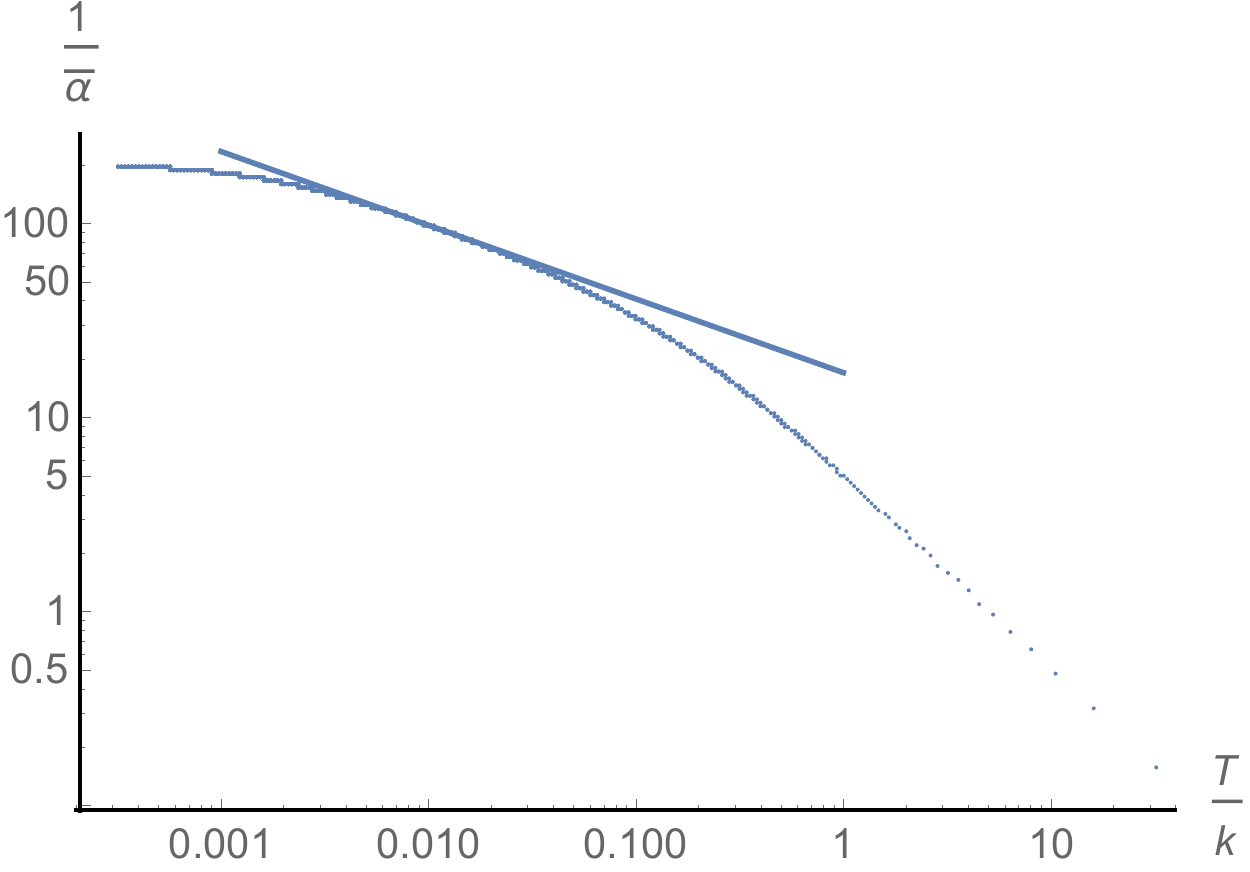}
\caption{ Plot for the conductivities $\sigma $ and $\bar\alpha$ as functions of the reduced temperature $T/k$ for a family of solutions with fixed $\mu/k=0.01$. The straight lines correspond to fits in the intermediate scaling regime. \label{fig3}}
\end{center}  
\end{figure}

In Figure \ref{fig3} we show these conductivities for the family of solutions presented in Figure \ref{fig2}.
From the plot we see that at high temperatures we have $\sigma/k\sim\bar\alpha\sim T/k$, which agrees with the expectations for a $CFT_4$. In the low temperature regime we have $\sigma\approx (T/k)^{0.14}$  and $\bar\alpha\approx 0.005$. Interestingly, for intermediate temperatures we see that the conductivities develop a power law scaling, in agreement with the intermediate scaling discussed previously. To be precise we find that for intermediate temperatures we have $\sigma/k\sim\left(T/k\right)^{1.88}$ and $\bar\alpha\sim \left(T/k \right)^{0.39}$.

\subsubsection{Computing $\alpha$ and $\bar\kappa$}

Now we move to compute the remaining conductivities, the thermoelectric coefficient $\alpha$ and the thermal conductivity $\bar\kappa$. In order to do that we implement the following perturbations to the metric and gauge field

\bea
\delta A&=& (-t \delta f_1(r)+\delta a_x(r))dx\,,\nn\\
\delta ds^2&=& (-t \delta f(r)+h_{tx}(r))dxdt + h_{rx}(r)dxdr.
\eea

Following as in the previous section we obtain expressions for $J$ and $Q$ that do not depend on the radial direction
\bea 
J&=&\frac{r h^4 \left(h_{tx} A_t'+t \delta f A_t'+2 r^2 f^2 g \left(a_x'-t \delta
  f_1'\right)\right)}{2 f},\nn\\
Q&=&-J A_t-\frac{1}{4}r^5 h^4f^3g^2\left[\left(\frac{h_{tx}}{r^2f^2g}\right)'+t\left(\frac{\delta f}{r^2f^2g}\right)'\right].
\eea

We can erase the temporal dependence fixing
\bea
\delta f_1(r)&=&E+\zeta A_t\,,\nn\\
\delta f(r)&=&2\zeta r^2f^2 g\,.
\eea

The remaining Einstein equation reads
\be
\delta h_{rx}=\frac{r \left(2 \zeta  f^2 \left(3 r^2 g\,h'^2+6 r \,g \,\,h h'+6 \left(r^2-g\right) h^2-k^2 h^6\right)-r h^2
   A_t' \left(\zeta  r A_t'+6 \zeta  A_t+6 E \right)\right)}{6 k^2 f^2 g \,h^6}\,.
\ee

Using the Eddington-Filkenstein coordinates, the expansions \eqref{NHsolv} and asking regularity at the horizon we can express $J$ and $Q$ as

\bea
J&=&\frac{1}{12} r_h^2 \left(\frac{a_{t_1} r_h \left(\zeta  r_h \left(a_{t_1}^2-12 f_0^2\right)+6 a_{t_1} E
   \right)}{f_0^2 k^2}+2 h_0^4 (a_{t_1} \zeta +6 r_h E )\right)\,,\nn\\
Q&=&\frac{r_h \left(r_h^2 \left(12-\frac{a_{t_1}^2}{f_0^2}\right)-2 h_0^4 k^2\right) \left(\zeta  r_h^2 \left(a_{t_1}^2-12
   f_0^2\right)+6 a_{t_1} r_h E +2 \zeta f_0^2 h_0^4 k^2\right)}{72 k^2}\,.
\eea

The transport coefficients then read

\bea
\alpha &=&\frac{\pi  a_{t_1} r_h^3}{f_0 k^2},\nn\\
\bar\kappa &=&-\frac{\pi  r_h^2 \left(a_{t_1}^2 r_h^2+2 f_0^2 \left(h_0^4 k^2-6 r_h^2\right)\right)}{6 f_0 k^2}
\eea

We see that $\alpha=\bar\alpha$ as it should be because the transport matrix is symmetric. This is a non trivial check on our computations. A quantity of interest is the thermal conductivity at zero electric current, which reads

\be 
\kappa=\bar\kappa-\frac{\alpha\bar\alpha T}{\sigma}=-\frac{\pi  f_0 h_0^4 r_h^2 \left(a_{t_1}^2 r_h^2+2 f_0^2 \left(h_0^4 k^2-6 r_h^2\right)\right)}{3
   \left(a_{t_1}^2+2 f_0^2 h_0^4 k^2\right)}.
\ee

\begin{figure}[!htb]
\begin{center}  
\includegraphics[scale=0.7]{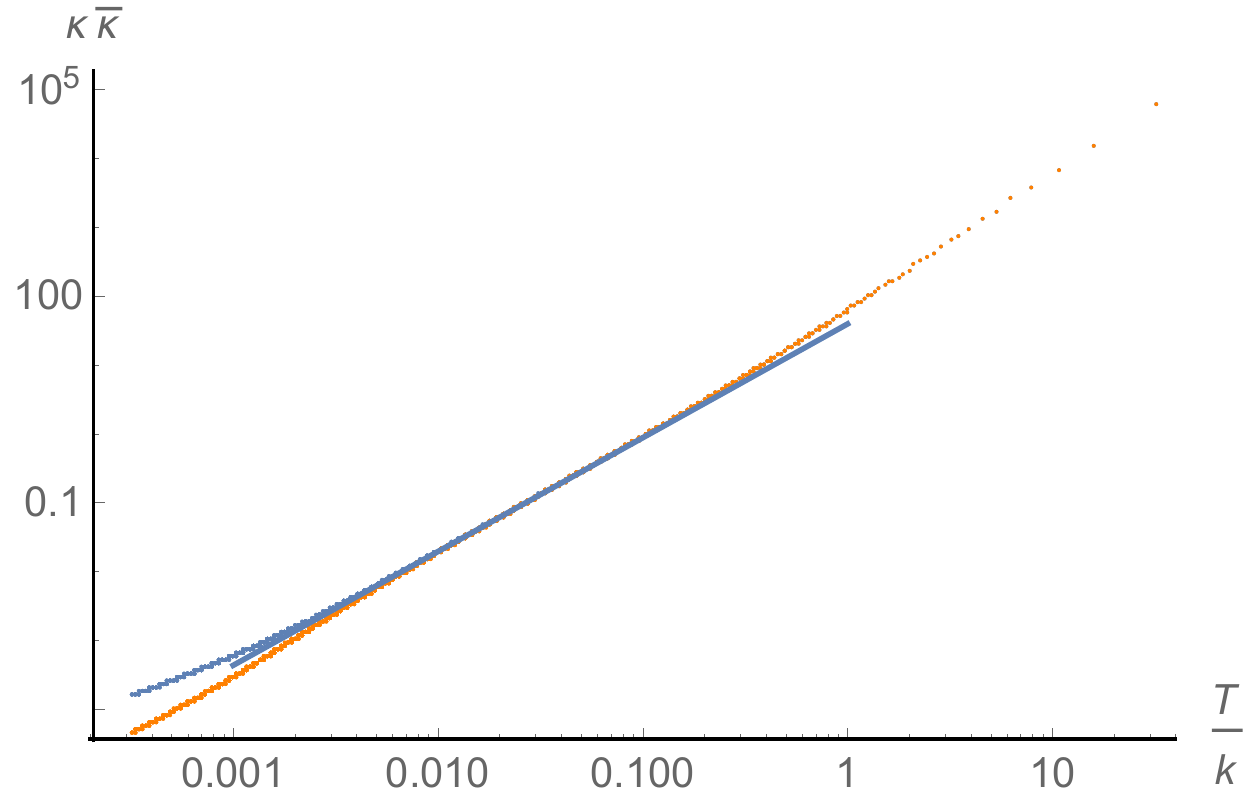}
\caption{  Plot for the conductivities $\bar\kappa $ (blue) and $\kappa$ (orange) as functions of the reduced temperature $T/k$ for a family of solutions with fixed $\mu/k=0.01$. The straight lines correspond to fits in the intermediate scaling regime. \label{fig4}}
\end{center}  
\end{figure}

In  Figure \ref{fig4} we plot the thermal conductivities $\bar\kappa$ and $\kappa$ for a family of black holes at fixed chemical potential $\mu/k=0.01$. For large temperatures we see that both these quantities scale as $\bar\kappa\sim\kappa\sim\left(T/k \right)^{2}$. For low temperatures, on the other hand, we have that $\bar\kappa\sim T/k$ while $\kappa\sim \left(T/k \right)^{3/2}$. In the intermediate regime we find that $\bar\kappa\sim\kappa\sim\left(T/k \right)^{1.66}$.

Dramatical differences between these two conductivities are indicators of the breakdown of the quasiparticle picture \cite{Mahajan:2013cja} .Moreover, in systems that behave as Fermi liquids the ratio $\kappa/(\sigma T)$ is constant and has a value of $\pi^2/(3 (k_B/e)^2)$, with $e$ the electric charge and $k_B$ the Boltzmann constant. Deviations from this constant value tell us that we are in a strong coupling regime.
 As we can see, $\kappa$ and $\bar\kappa$ scale differently in the low temperature limit but get closer together as we increase the temperature.

\section{Nil black holes with intermediate scaling}
\label{sec3}

We will now consider charged black holes solutions with Nil horizon geometry. 
  
 \subsection{Solutions}
\label{sec3a}

Let us then consider the ansatz 
\bea
A&=&A_t(r)\  dt\, , \label{ansatznil} \nn\\ 
ds^2&=&- r^{2} g(r)f(r)^2 dt^2 + \frac1{r^2 g(r)} dr^2 +r^2 h^2(r)\left( dx^2 +  dy^2\right)+ \frac{r^2}{h^4(r)}\left(   dz-kx dy   \right)^ 2\,,
\eea

for which the equations of motion read
\bea
\frac{2 r^2 \left(12-\frac{A_t'^2}{f^2}\right)-\frac{k^2}{h^8}}{6 r^3}-g'+g \left(-\frac{2 r h'^2}{h^2}-\frac{4}{r}\right)&=&0\,, \nn\\
h''-\frac{\frac{2 r^3 h' \left(A_t'(r)^2-3 f^2 (g+4)\right)}{f^2}+\frac{6 r^4 g h'^2}{h}+\frac{k^2 r
   h'}{h^8}-\frac{2 k^2}{h^7}}{6 r^4 g}&=&0\,,\nn\\
f'-\frac{2 r f h'^2}{h^2}&=&0\,,\nn\\
A_t''-\frac{A_t' \left(r f'-3 f\right)}{r f}&=&0\,.
\eea

We will look for black hole solutions to these equations by integrating out the fields from the near horizon
\bea
A_t(r)&\simeq & a_{t_1}(r-r_h)-\frac{a_{t_1} \left(12 h_0^{16} r_h^4 \left(a_{t_1}^2-12 f_0^2\right)^2+12 f_0^2 h_0^8 k^2 r_h^2 \left(a_{t_1}^2-12
   f_0^2\right)-5 f_0^4 k^4\right)}{2 r_h \left(2 h_0^8 r_h^2 \left(a_{t_1}^2-12 f_0^2\right)+f_0^2 k^2\right)^2}(r-r_h)^2+\ldots, \nn \\
f(r)&\simeq & f_0+\frac{8 f_0^5 k^4}{r_h \left(2 h_0^8 r_h^2 \left(a_{t_1}^2-12 f_0^2\right)+f_0^2 k^2\right)^2}(r-r_h)+\ldots, \nn \\
g(r)&\simeq &\frac{r_h^2 \left(24-\frac{2 a_{t_1}^2}{f_0^2}\right)-\frac{k^2}{h_0^8}}{6 r_h^3}(r-r_h)+\ldots, \nn \\
h(r)&\simeq & h_0+\frac{2 f_0^2 h_0 k^2}{2 h_0^8 r_h^3 \left(a_{t_1}^2-12 f_0^2\right)+f_0^2 k^2 r_h}(r-r_h)+\ldots,
\eea
towards the boundary
\bea
A_t(r)&\simeq & \mu+\frac\rho{r^2}-\frac{\rho \,k^4}{216 h_\infty^{16}r^6}+ \ldots,  \nn\\ 
f(r)&\simeq & f_\infty-\frac{f_\infty k^4}{72h_\infty^{16}r^4}+\ldots, \nn\\
g(r)&\simeq &1 -\frac{k^2}{12h_\infty^8 r^2}+\frac{g_4^\infty}{r^4}+\frac{k^4\log r}{18 h_\infty^{16}r^4}   +\ldots,\nn\\
h(r)&\simeq & h_\infty+\frac{k^2}{12h_\infty^7 r^2}+\frac{h_4^\infty}{r^4}-\frac{k^4 \log r}{18 h_\infty^{15} r^4}+\ldots,
\label{uvnil}
\eea
where we will shoot to find solutions with $f_\infty=h_\infty=1$.
Again, this boundary condition imply that our black hole solutions have the same scaling for the metric (in all the directions) when we move towards the boundary. The thermodynamics of the black hole are given by
\bea
T&=&\frac{\frac{a_{t_1}^2 r_h}{3 f_0}+\frac{f_0 k^2}{6 h_0^8 r_h}-4 f_0 r_h}{4 \pi },\\
s&=&2\pi A_h=2\pi r_h^3.
\eea
As in the previous section, we will characterize the family of solutions with the dimensionless parameters $T/k$ and $\mu/k$.

\begin{figure}[!htb]
\begin{center}  
\includegraphics[scale=0.63]{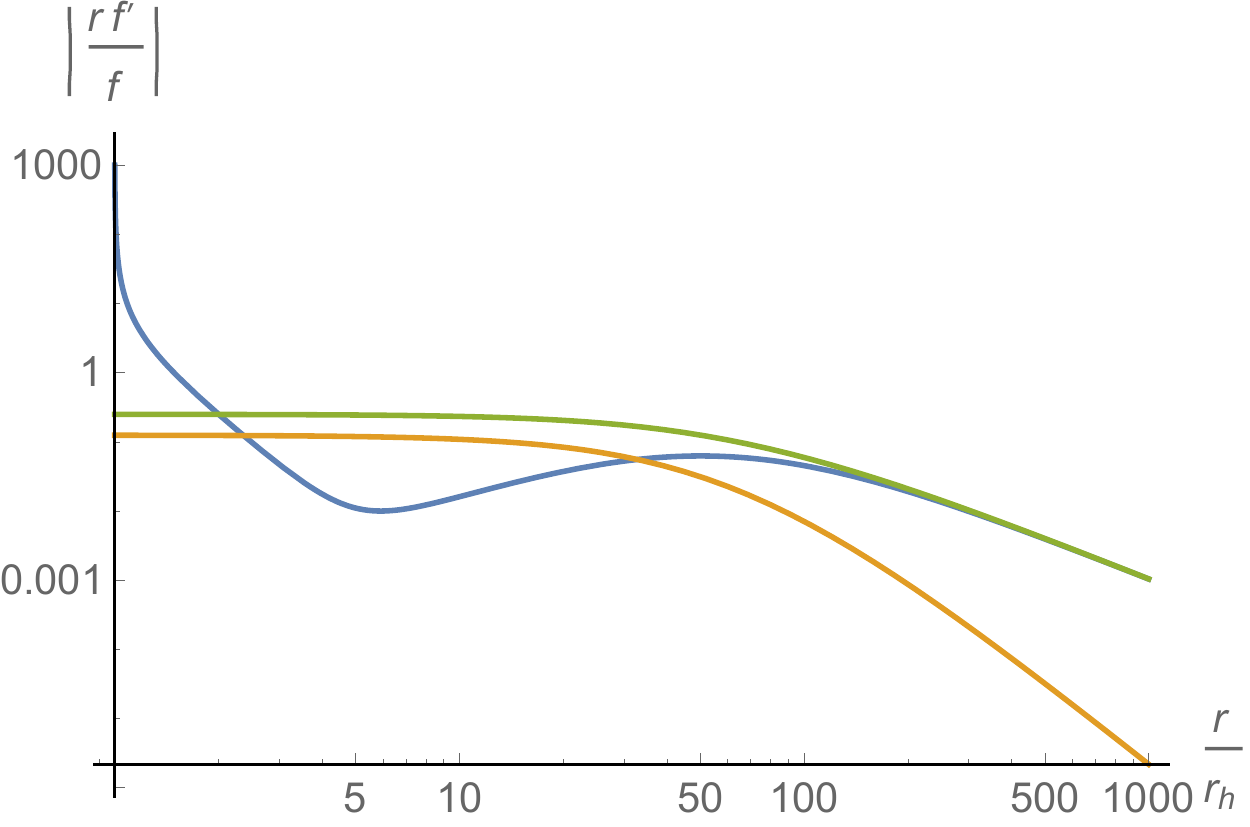}\hfill\includegraphics[scale=0.63]{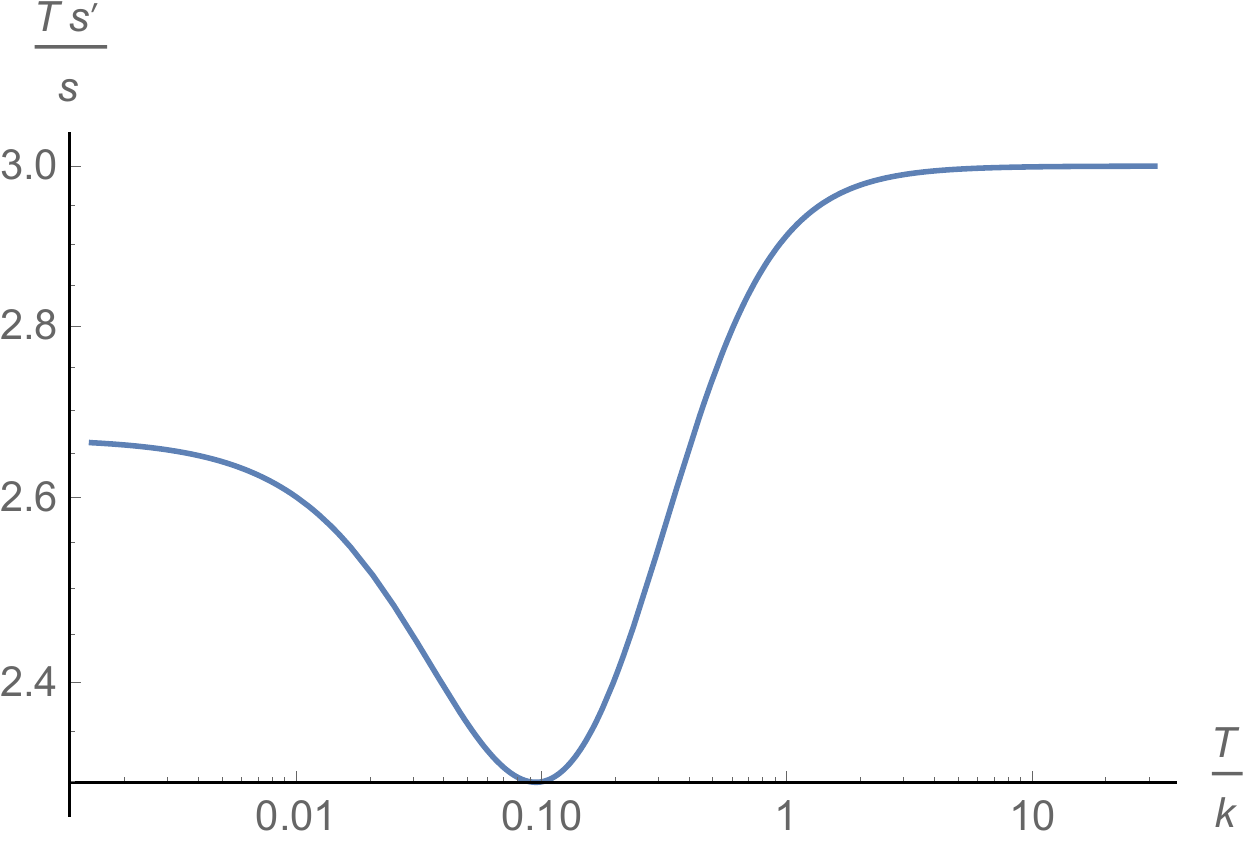}
\caption{Left: Typical profile for $f(r)$ (blue), $g(r)$ (orange) and $h(r)$ (green). The combination $rf'/f$, where $f$ corresponds to any of the aforementioned functions, was chosen to make manifest the emergence of new scalings towards the IR. The plot corresponds to a solution with $T/k=0.00147$ and $\mu/k=0$. Right: $T s'/s$ as a function of the reduced temperature $T/k$.  \label{fig1n}}
\end{center}  
\end{figure}

For simplicity, let us begin by studying solutions with $\mu=0$, corresponding to neutral black holes.
Integrating the equations of motion we find a family of solutions with different $T/k$. 
On the right panel of 
Figure \ref{fig1n} we show how the entropy of the corresponding solutions scales with respect to the temperature. Again, for high enough temperatures we see the expected $CFT_4$ related scaling $s/k^3\sim (T/k)^3$. On the other hand,
for low enough temperatures we find a new scaling $s/k^3\sim (T/k)^{30/11}$. To understand better the nature of this new scaling, we must study in detail the behavior of the metric fields.

In the left hand side of Figure \ref{fig1n} we show the scalings related to the profiles of the metric functions for a low $T/k$ solution. Far away from the horizon, we see that the metric behave as demanded in (\ref{uvnil}). On the other hand, we find that near the horizon the metric functions behave 
 differently $h\sim r^{-1/4}$ and $g\sim r^{1/8}$.
 Plugging these scalings back into the metric give quite a peculiar scaling
 \bea
t \rightarrow \lambda^{11/8}\, t\,,\,\,\,\, \omega_1\rightarrow \lambda\, \omega_1\, , \,\,\,\, \omega_2\rightarrow \lambda\, \omega_2 \,, \,\,\,\, \omega_3\rightarrow \lambda^{3/2} \omega_3\,,\label{scalnil}
\eea
which gives the correct scaling for the entropy as a function of the temperature  \cite{Donos:2014oha}.
 Unfortunately we do not know of a deformed theory from which we can extract exactly this kind of solutions.
 
\begin{figure}[!htb]
\begin{center}  
\includegraphics[scale=0.63]{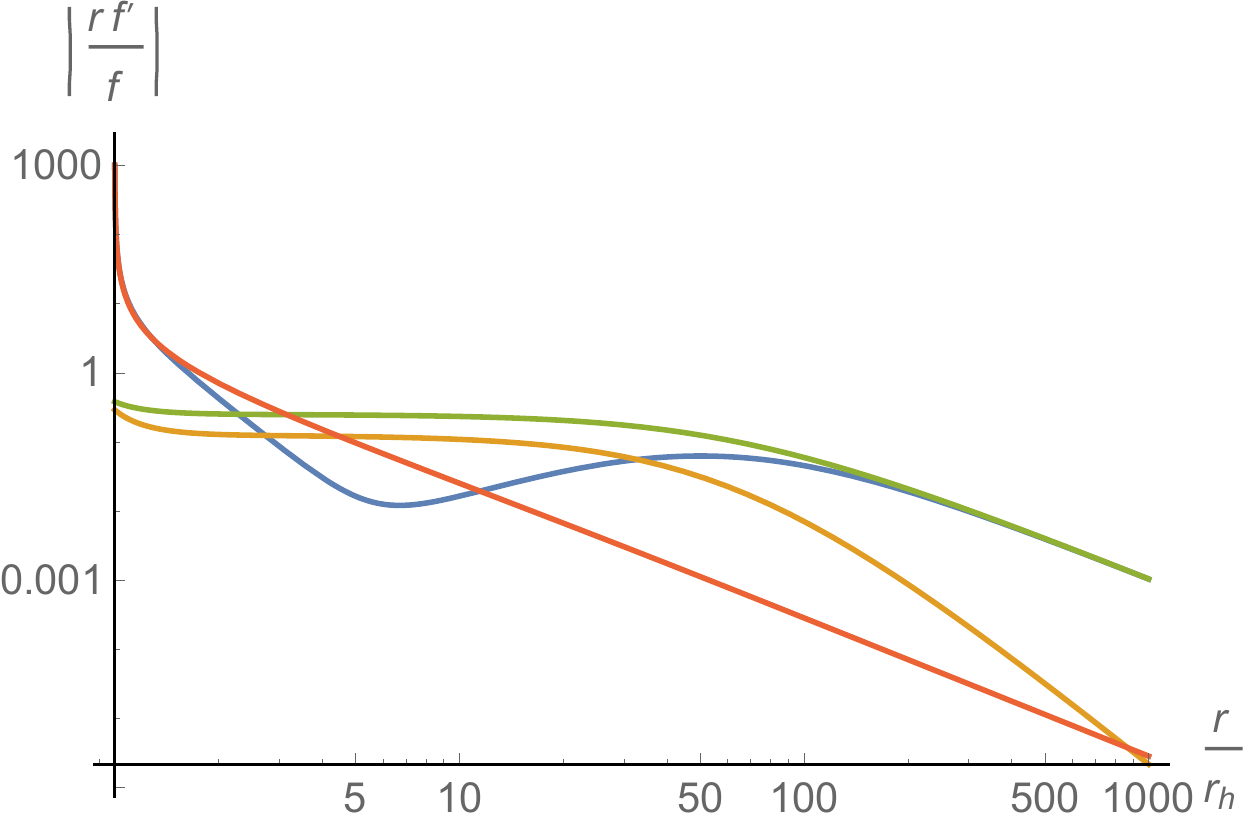}\hfill\includegraphics[scale=0.63]{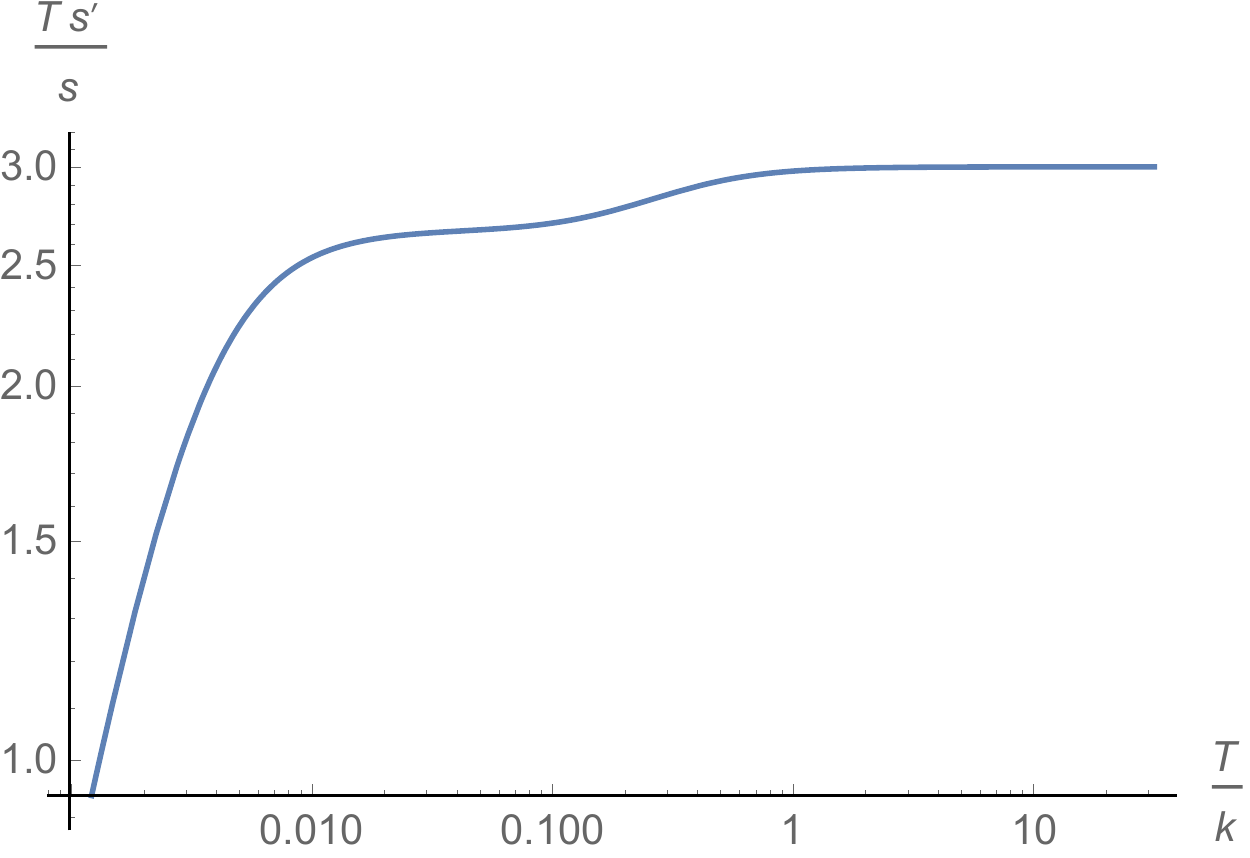}
\caption{Left: Typical profile for $f(r)$ (blue), $g(r)$ (orange), $h(r)$ (green) and $A_t(r)$ (red). The combination $rf'/f$, where $f$ corresponds to any of the aforementioned functions, was chosen to make manifest the emergence of new scalings towards the IR. The plot corresponds to a solution with $T/k=0.0003185$ and $\mu/k=0.01$. Right: $T s'/s$ as a function of the reduced temperature $T/k$.  \label{fig2s}}
\end{center}  
\end{figure}
 
 Let us now study the charged solutions.
 In Figure \ref{fig2s} we show typical profiles for the fields and the scaling of the entropy for a family of solutions with fixed $\mu/k=0.01$. Again the neutral IR scaling becomes an intermediate scaling and the charged solutions flow to $AdS_2\times Nil$ at the IR.

 \subsection{Finite conductivities from Nil horizons}
\label{sec3b}

In this section we will compute the transport coefficients for this Einstein-Maxwell solutions and shows how they scale along the RG flow. In order to express the transport coefficients as functions of horizon data we must follow the same steps we used in the previous section, then we are not going to repeat the procedure and we are just going to highlight the final results.

In this case the fluctuations must be along the $x-$ direction,
\bea
\delta A&=&(-E t+\delta a_x(r))dx\,,\nn\\
\delta ds^2&=&h_{tx}(r)dtdx+h_{rx}(r)drdx\,,
\eea
with the constant $E$ the applied (DC) electric field. From the Einstein-Maxwell equations we compute the conserved current and charge $J$ and $Q$ and the remaining equation for $h_{rx}$ reads

\bea
h_{rx}&=&-\frac{4 E  h^8 A_t'}{k^2 f^2 g}\,.\label{eomhrxnil}
\eea

Using the near horizon data and asking for regularity we obtain

\bea 
J&=&\frac{2 a_{t_1}^2 h_0^6 r_h^3 E }{f_0^2 k^2}-\frac{r_h E }{h_0^2}\,,\nn\\
Q&=&-\frac{a_{t_1} h_0^6 r_h^2 E  \left(r_h^2 \left(24-\frac{2 a_{t_1}^2}{f_0^2}\right)-\frac{k^2}{h_0^8}\right)}{6 k^2}\,,
\eea
and from this we can compute the conductivities $\sigma$ and $\bar\alpha$ as
\bea
\sigma&=&\frac{\partial J}{\partial E}=\frac{2 a_{t_1}^2 h_0^6 r_h^3}{f_0^2 k^2}+\frac{r_h}{h_0^2}\,,\nn \\
\bar\alpha&=&\frac{1}{T}\frac{\partial Q}{\partial E}=\frac{4 \pi  a_{t_1} h_0^6 r_h^3}{f_0 k^2}\,.
\eea

\begin{figure}[!htb]
\begin{center}  
\includegraphics[scale=0.63]{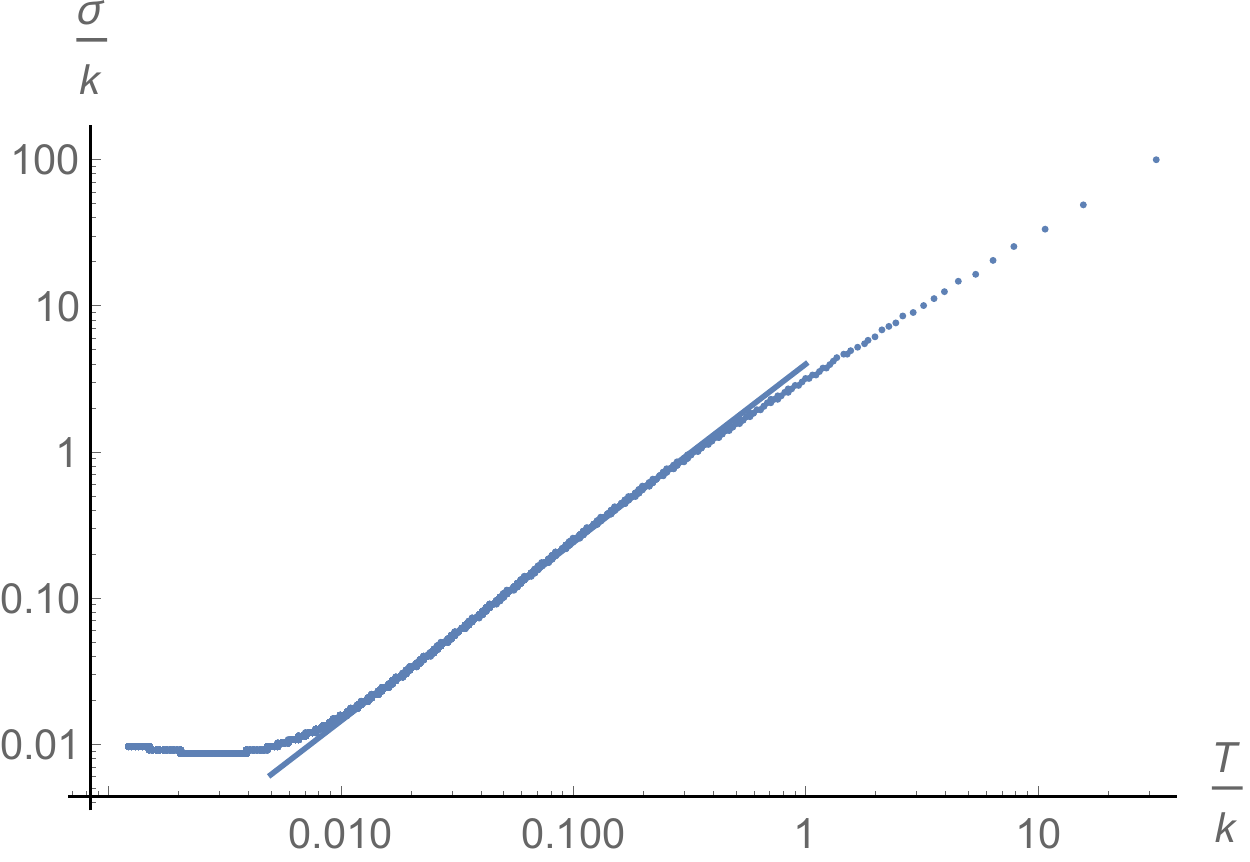}\hfill\includegraphics[scale=0.63]{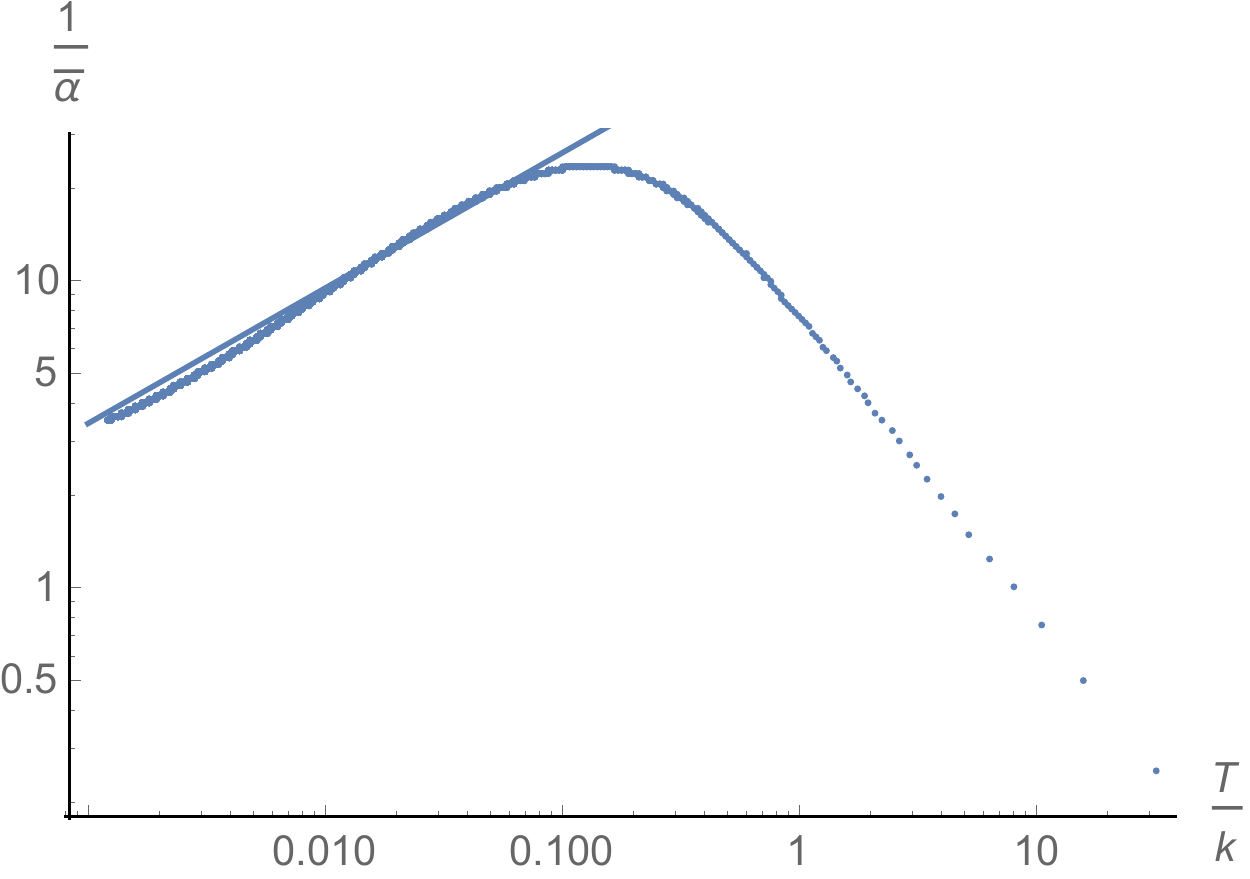}
\caption{ Plot for the conductivities $\sigma $ and $\bar\alpha$ as functions of the reduced temperature $T/k$ for a family of solutions with fixed $\mu/k=0.01$. The straight lines correspond to fits in the intermediate scaling regime. \label{fig3s}}
\end{center}  
\end{figure}

In Figure \ref{fig3s} we show these conductivities for the family of solutions presented in Figure \ref{fig2s}.
From the plot we see that the conductivities show that at high temperatures we have that $\sigma\sim\bar\alpha\sim T/k$ which agrees with the expectations for a $CFT_4$. In the low temperature regime we have $\sigma\sim 0.22$ and $\bar\alpha\sim 0.44$. Interestingly, form intermediate temperatures we see that the conductivities develop a power law scaling $\sigma\sim\left(T/k\right)^{1.38}$ and $\bar\alpha\sim \left(T/k \right)^{-0.44}$. Another interesting feature of the conductivity $\sigma$ is that it is not monotonous with the temperature, giving a metalic behaviour at low $T/k$ while an insulator one at high temperatures \cite{Baggioli:2015zoa}.

Now to compute the remaining conductivities we perturb the Einstein-Maxwell solutions as

\bea
\delta A&=& (-t \delta f_1(r)+\delta a_x(r))dx\,,\nn\\
\delta ds^2&=& (t \delta f(r)+h_{tx}(r))dxdt + h_{rx}dxdr\,.
\eea

Following as in the previous section we obtain expressions for $J$ and $Q$ that does not depend on the radial direction and can be just expressed as function of IR data

\bea
J&=&\frac{r_h \left(\frac{2 a_{t_1} h_0^8 r_h^2 \left(\zeta r_h \left(a_{t_1}^2-12 f_0^2\right)+6 a_{t_1} E
   \right)}{f_0^2 k^2}+a_{t_1} \zeta  r_h+6 E \right)}{6 h_0^2}\,,\nn\\
Q&=&-\frac{r_h \left(2 h_0^8 r_h^2 \left(a_{t_1}^2-12 f_0^2\right)+f_0^2 k^2\right) \left(2 h_0^8 r_h \left(\zeta 
   r_h \left(a_{t_1}^2-12 f_0^2\right)+6 a_{t_1} E \right)+\zeta f_0^2 k^2\right)}{72 f_0^2 h_0^{10} k^2}\,.
\eea

Where we erased the temporal dependence fixing
\bea
\delta f_1(r)&=&E+\zeta A_t(r)\,,\nn\\
\delta f(r)&=&2\zeta r^2f(r)^2 g(r)\,,
\eea
and we use that 
\be
\delta h_{rx}=-\frac{2 r h^8 A_t' \left(\zeta  r A_t'+6 \zeta  A_t+6 E \right)+\zeta  f^2 \left(12 r^2 h^6 \left(-r^2 g
   h'^2+r g h h'+2 (g-1) h^2\right)+k^2\right)}{3 k^2 r f^2 g}\,,
\ee

Then, the remaining transport coefficients read

\bea
\alpha &=&\frac{4 \pi  a_{t_1} h_0^6 r_h^3}{f_0 k^2},\nn\\
\bar\kappa &=&-\frac{\pi  r_h^2 \left(\frac{2 h_0^8 r_h^2 \left(a_{t_1}^2-12 f_0^2\right)}{k^2}+f_0^2\right)}{3 f_0 h_0^2};
\eea

Again we obtain the expected result $\alpha=\bar\alpha$. The thermal conductivity at zero electric current is written in this case as

\be 
\kappa=\bar\kappa-\frac{\alpha\bar\alpha T}{\sigma}=-\frac{2 \pi  f_0 h_0^8 r_h^4 \left(a_{t_1}^2-12 f_0^2\right)+\pi  f_0^3 k^2 r_h^2}{3 h_0^2 \left(2 a_{t_1}^2
   h_0^8 r_h^2+f_0^2 k^2\right)}.
\ee

\begin{figure}[!htb]
\begin{center}  
\includegraphics[scale=0.7]{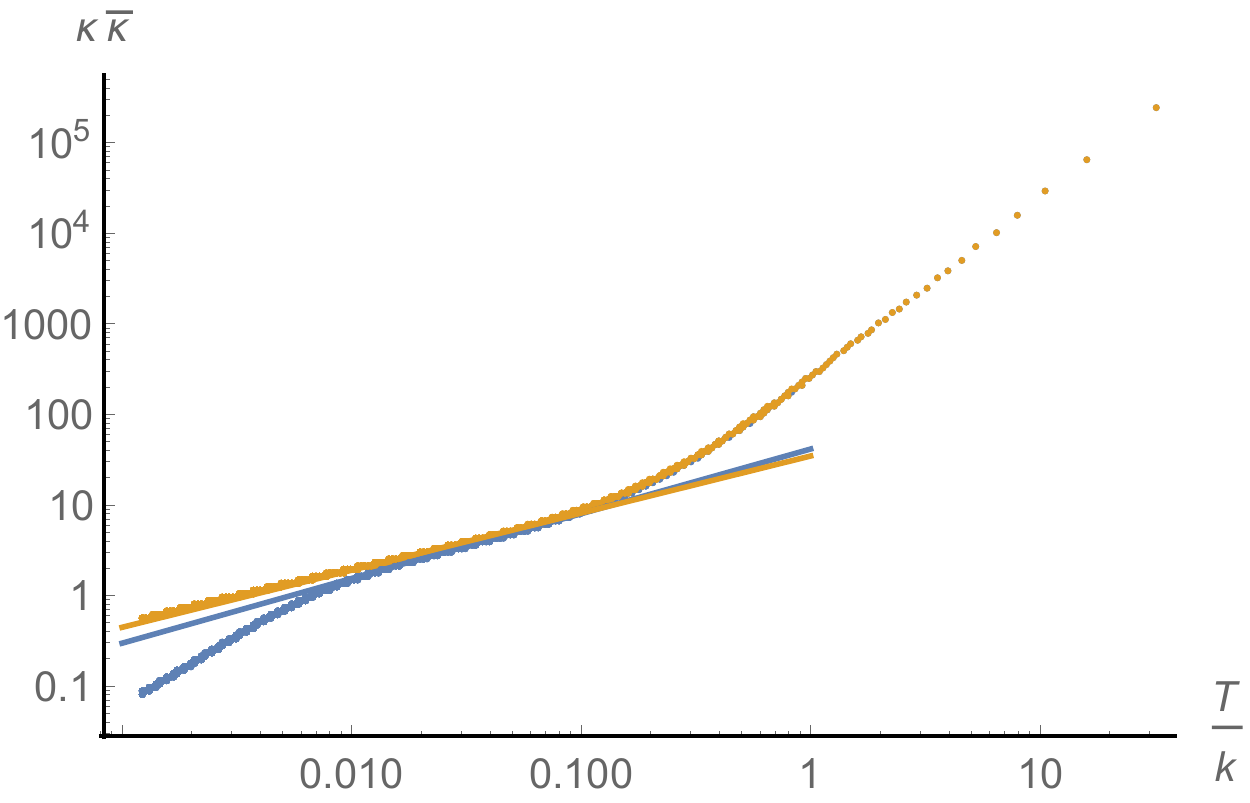}
\caption{ Conductivities $\bar\kappa $ (blue) and $\kappa$ (orange) as functions of the reduced temperature $T/k$ for a family of solutions with fixed $\mu/k=0.01$. \label{fig4s}}
\end{center}  
\end{figure}

In the left panel of Figure \ref{fig4s} we plot the thermal conductivities $\kappa$ and $\bar\kappa$  for a family of black holes at fixed chemical potential $\mu/k=0.01$. For large temperatures we see that $\bar\kappa\sim\left(T/k \right)^{2}$. For low temperatures we have that $\bar\kappa\sim 0.54$ while  $\kappa\sim \left(T/k \right)^{1.45}$ . In the intermediate regime we find that $\bar\kappa\sim\left(T/k \right)^{0.63}$ while  $\kappa\sim\left(T/k \right)^{0.71}$. Again, the qualitative difference between $\kappa$ and $\bar\kappa$ signal the break of the quasi-particle picture at low enough temperatures.

\newpage

\section{$SL_2(\cal R)$ black holes with intermediate scaling}
\label{sec4}

Finally, let us consider black brane solutions corresponding to foliations of an $SL_2(\cal R)$ metric.

\subsection{Solutions}

We will work with the ansatz
\bea
A&=&A_t(r)\  dt\, , \label{ansatzsl2r} \nn\\ 
ds^2&=&- r^{2} g(r)f(r)^2 dt^2 + \frac1{r^2 g(r)} dr^2 +\frac{r^2 h^2(r)}{k^2x^2}\left( dx^2 +  dy^2\right)+ \frac{r^2}{h^4(r)}\left(   dz+ \frac{dy}{kx}   \right)^ 2\,.
\eea

The equations of motion read
\bea
\frac{2 r^2 \left(\frac{A_t'^2}{f^2}-12\right)+12 g \left(\frac{r^2 h'^2}{h^2}+1\right)+\frac{(4 h^6+1)}{ k^2 h^8}}{6 r}+g'&=&0,\nn\\
h''-\frac{2 r^3 h^8 A_t'^2 h'+6 r^4 f^2 g h^7 h'^2-6 r^3 f^2 g h^8 h'+4 k^2 r f^2 h^6 h'+k^2 r f^2
   h'-24 r^3 f^2 h^8 h'-2 k^2 f^2 h^7-2 k^2 f^2 h}{6 r^4 f^2 g h^8}&=&0,\nn\\
f'-\frac{2 r f h'^2}{h^2}=0,\,\,\,\,\,\,\,\,\,\,\,A_t''-\frac{A_t' \left(r f'-3 f\right)}{r f}=0.
\eea

\begin{figure}[!h]
\begin{center}  
\includegraphics[scale=0.63]{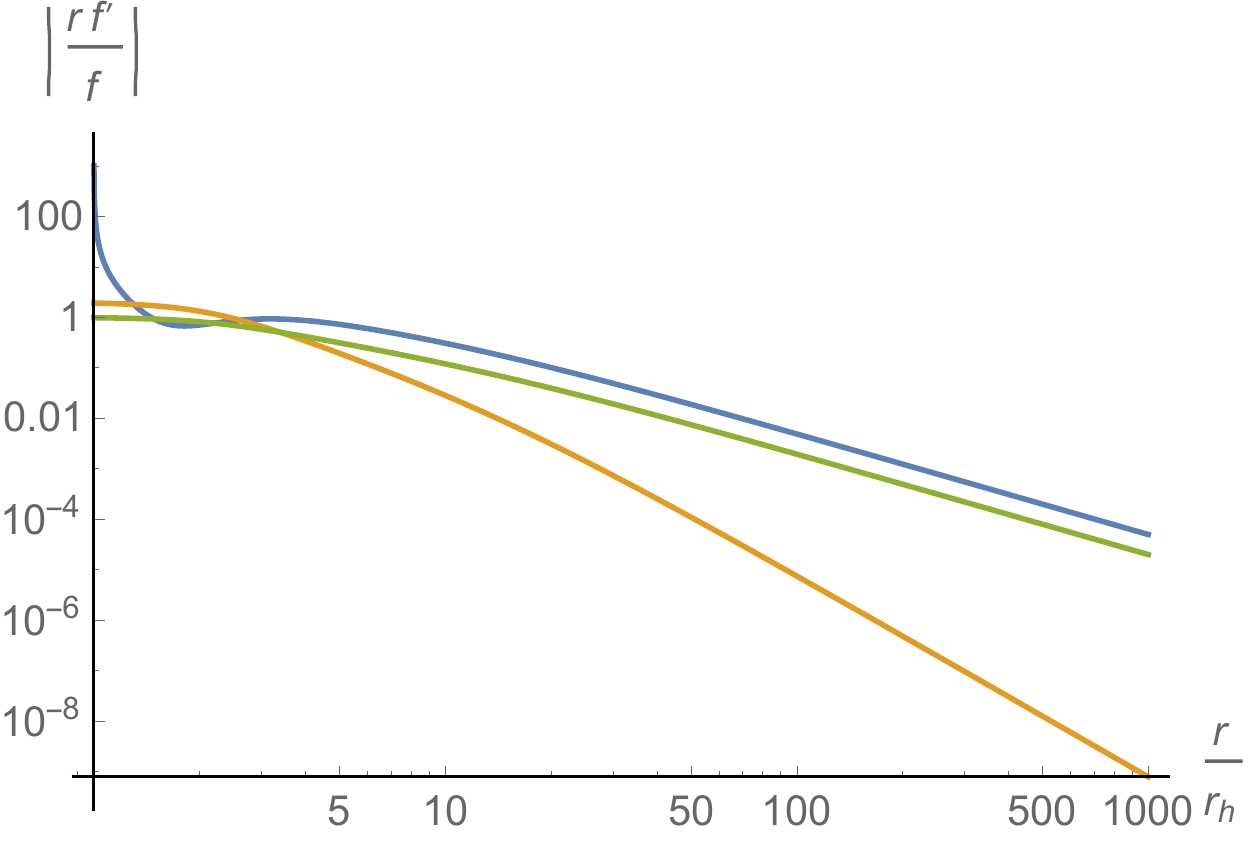}\hfill\includegraphics[scale=0.63]{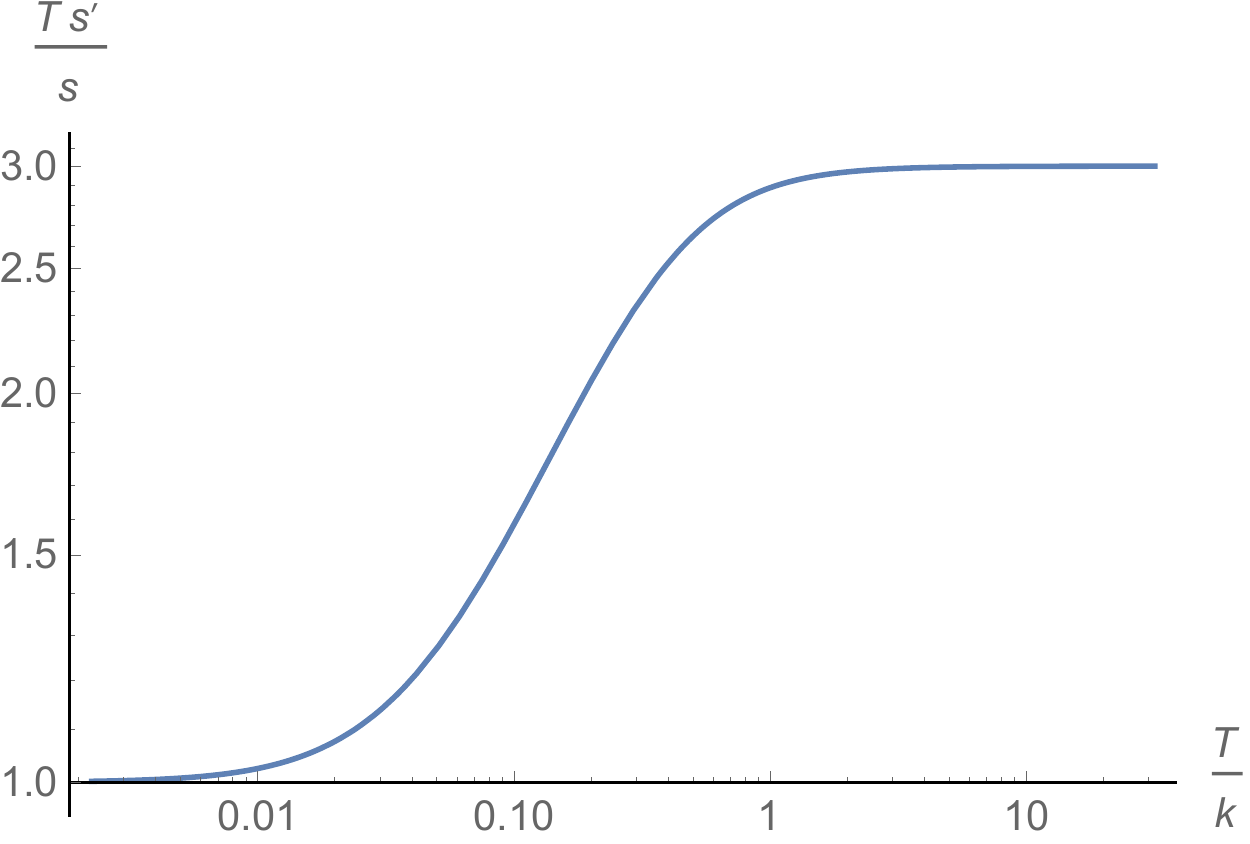}
\caption{Left: Typical profile for $f(r)$ (blue), $g(r)$ (orange) and $h(r)$ (green). The combination $rf'/f$, where $f$ corresponds to any of the aforementioned functions, was chosen to make manifest the emergence of new scalings towards the IR. The plot corresponds to a solution with $T/k=0.00225$ and $\mu/k=0$. Right: $T s'/s$ as a function of the reduced temperature $T/k$.  \label{fig1sl2r}}
\end{center}  
\end{figure}

We will integrate numerically these equations from the near horizon
\bea
A_t(r)&\simeq & a_{t_1}(r-r_h)-\frac12a_{t_1}   \left(3- \frac{f_1}{f_0} \right) (r-r_h)^2
+ \ldots,  \nn\\ 
f(r)&\simeq & f_0+\frac{8 f_0^5 \left(h_0^6+1\right)^2 k^4}{r_h \left(2 h_0^8 r_h^2 \left(a_{t_1}^2-12 f_0^2\right)+f_0^2
   \left(4 h_0^6+1\right) k^2\right)^2}(r-r_h)+\ldots, \nn\\
g(r)&\simeq &\left(\frac{r_h^2 \left(24-\frac{2 a_{t_1}^2}{f_0^2}\right)-\frac{\left(4 h_0^6+1\right) k^2}{h_0^8}}{6 r_h^3}\right)(r-r_h) + \ldots,\nn\\
h(r)&\simeq & h_0+\frac{2 f_0^2 h_0\left(h_0^6+1\right) k^2}{2 h_0^8 r_h^3 \left(a_{t_1}^2-12 f_0^2\right)+f_0^2 \left(4
h_0^6+1\right) k^2 r_h}(r-r_h)+\ldots,\label{NHSl2R}
\eea
towards the boundary
\bea
A_t(r)&\simeq & \mu+\frac\rho{r^2}-\frac{\rho(1+h_\infty^6)^2k^4}{216h_\infty^{16}r^6}+ \dots,  \nn\\ 
f(r)&\simeq & f_\infty - \frac{f_\infty(1+h_\infty^6)^2 k^4}{72 h_\infty^{16}r^4}     +  \ldots, \nn\\
g(r)&\simeq &1-\frac{(1+4h_\infty^6)k^2}{12h_\infty^8r^2}+\frac{g^{\infty}_4}{r^4}+\frac{(1+h_\infty^6)^2k^4 \log r}{18h_\infty^{16}r^4}+\ldots,\nn\\
h(r)&\simeq & h_\infty+\frac{(1+h_\infty^6)k^2}{12 h_\infty^7r^2}+\frac{h^{\infty}_4}{r^4}+\ldots,
\eea
through a shooting method. A typical solution is shown in Figure \ref{fig1sl2r}.

The black hole thermodynamics is dictated by 
\bea
T&=&\frac{f_0 \left(r_h^2 \left(24-\frac{2 a_{t_1}^2}{f_0^2}\right)-\frac{\left(4 h_0^6+1\right) k^2}{h_0^8}\right)}{24 \pi 
 r_h}\,, \nn \\
s&=&2 \pi r_h^3 \,.
\eea

We are now ready to construct a family of solutions that is characterized by the parameter $T/k$ and $\mu/k$. Again we first look for solutions with $\mu=0$ and we find that for low enough $T/k$ the metric profiles develop a scaling in the deep IR near the black hole horizon.
This scaling is shown in  Figure \ref{fig1sl2r} where we see for a particular solution that near the horizon $f\sim (r/r_h)^2$ and $h\sim (r/r_h)^{-1}$. This implies that the dual field theory will have an IR scaling dictated by
 \bea
t \rightarrow \lambda^{1}\, t\,,\,\,\,\, \omega_1\rightarrow \, \omega_1\, , \,\,\,\, \omega_2\rightarrow\, \omega_2 \,, \,\,\,\, \omega_3\rightarrow \lambda^{1} \omega_3\,.\label{scalsl2r}
\eea
This scaling is also reflected on the fact that the entropy scales as $s/k^3\sim T/k$ at low temperatures.

\begin{figure}[!h]
\begin{center}  
\includegraphics[scale=0.63]{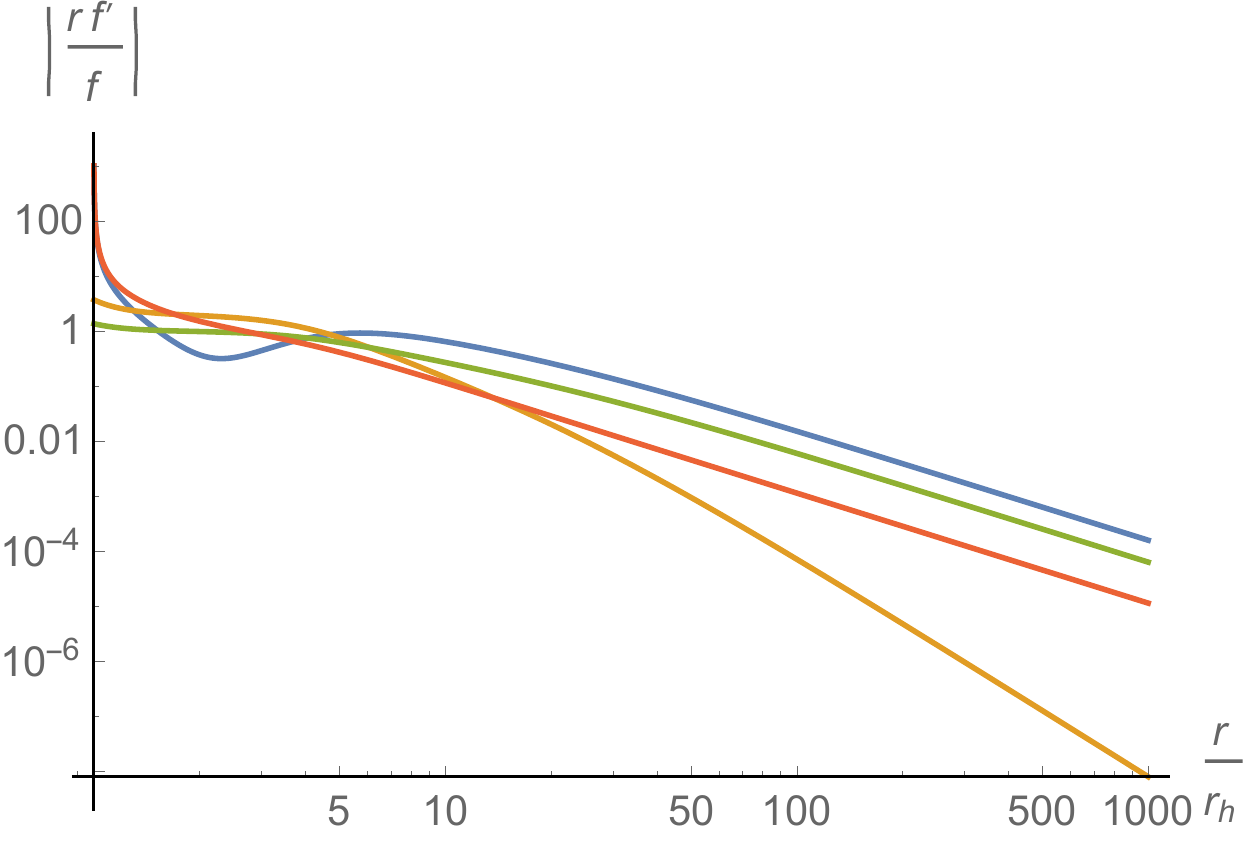}\hfill\includegraphics[scale=0.63]{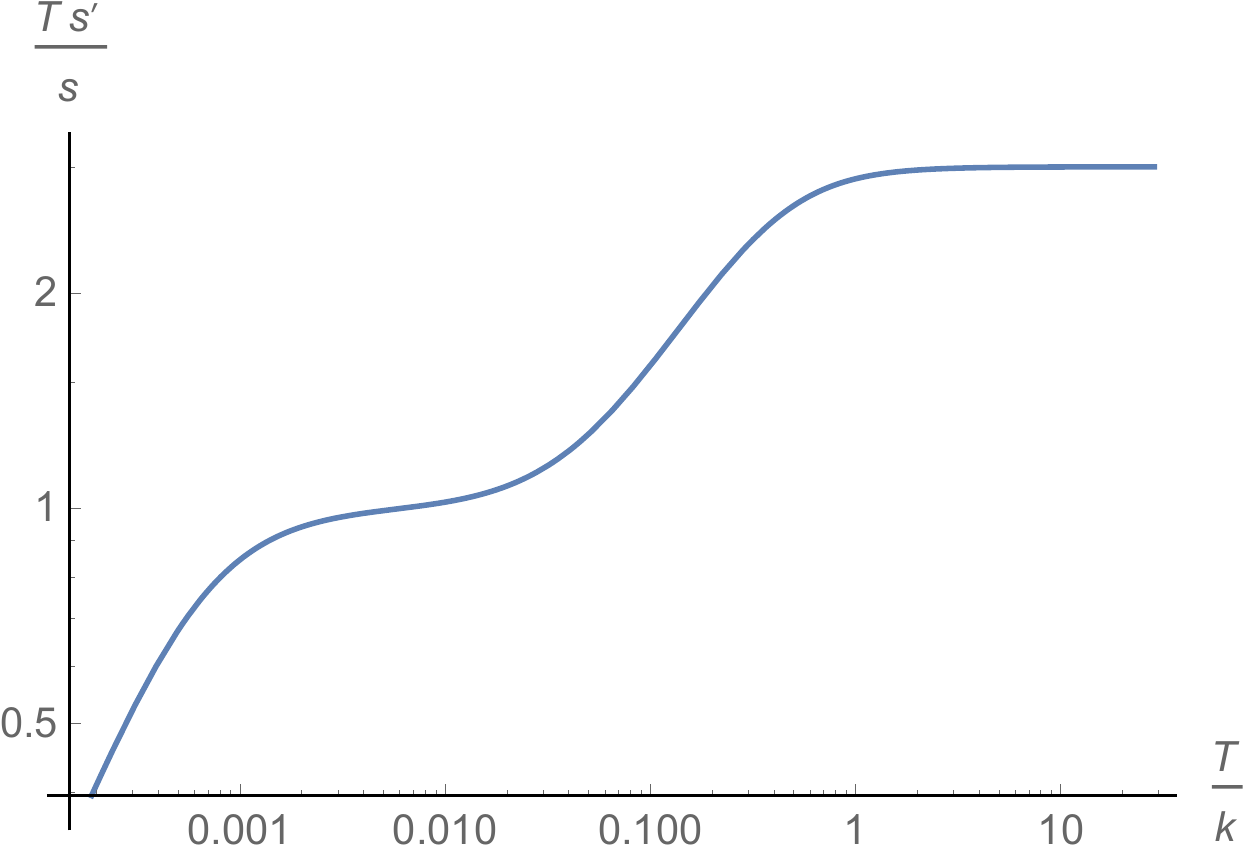}
\caption{Left: Typical profile for $f(r)$ (blue), $g(r)$ (orange), $h(r)$ (green) and $A_t(r)$ (red). The combination $rf'/f$, where $f$ corresponds to any of the aforementioned functions, was chosen to make manifest the emergence of new scalings towards the IR. The plot corresponds to a solution with $T/k=0.0001892$ and $\mu/k=0.01$. Right: $T s'/s$ as a function of the reduced temperature $T/k$.  \label{fig2sl2r}}
\end{center}  
\end{figure}

When we turn on the chemical potential, the IR scalings become intermediate scalings and the near horizon geometry becomes $AdS_2\times SL_2(\cal R)$. A typical family of solutions and profile is shown in Figure \ref{fig2sl2r}. In the following section we will use this family of solutions to study the DC transport porperties of the dual field theory.

\newpage

\subsection{Computing the transport coefficients}

Lets start with the fluctuations along the $x-$ direction,
\bea
\delta A&=&(-E t+\delta a_x(r))dx\nn\\
\delta ds^2&=&h_{tx}(r)dtdx+h_{rx}(r)drdx,
\eea
with  $E$ the constant electric electric field. From the Einstein and Maxwell equations we obtain the conserved current and charge $J$ and $Q$ and the remaining equation for $h_{rx}$ reads
\bea
h_{rx}&=&-\frac{4 E  h^8 A_t'}{k^2 f^2 g \left(2 h^6+1\right)}\,.\label{eomhrxsl2r}
\eea
Using the near horizon data \eqref{NHSl2R} and asking for regularity we have that the conserved quantities read
\bea 
J&=&\frac{2 a_{t_1}^2 h_0^6 r_h^3 E}{f_0^2 \left(2 h_0^6+1\right) k^2}+\frac{r_h E }{h_0^2}\,,\nn\\
Q&=&\frac{a_{t_1} h_0^6 r_h^2 E \left(r_h^2 \left(24-\frac{2 a_{t_1}^2}{f_0^2}\right)-\frac{\left(4 h_0^6+1\right)
   k^2}{h_0^8}\right)}{6 \left(2 h_0^6+1\right) k^2}\,.
\eea
From this we compute the conductivities $\sigma$ and $\bar\alpha$ which have the following expression
\bea
\sigma&=&\frac{\partial J}{\partial E}=\frac{2 a_{t_1}^2 h_0^6 r_h^3}{f_0^2 \left(2 h_0^6+1\right) k^2}+\frac{r_h}{h_0^2}\,,\nn\\
\bar\alpha&=&\frac{1}{T}\frac{\partial Q}{\partial E}=\frac{4 \pi  a_{t_1}h_0^6 r_h^3}{k^2 \left(2 f_0 h_0^6+f_0\right)}\,.
\eea

\begin{figure}[!htb]
\begin{center}  
\includegraphics[scale=0.63]{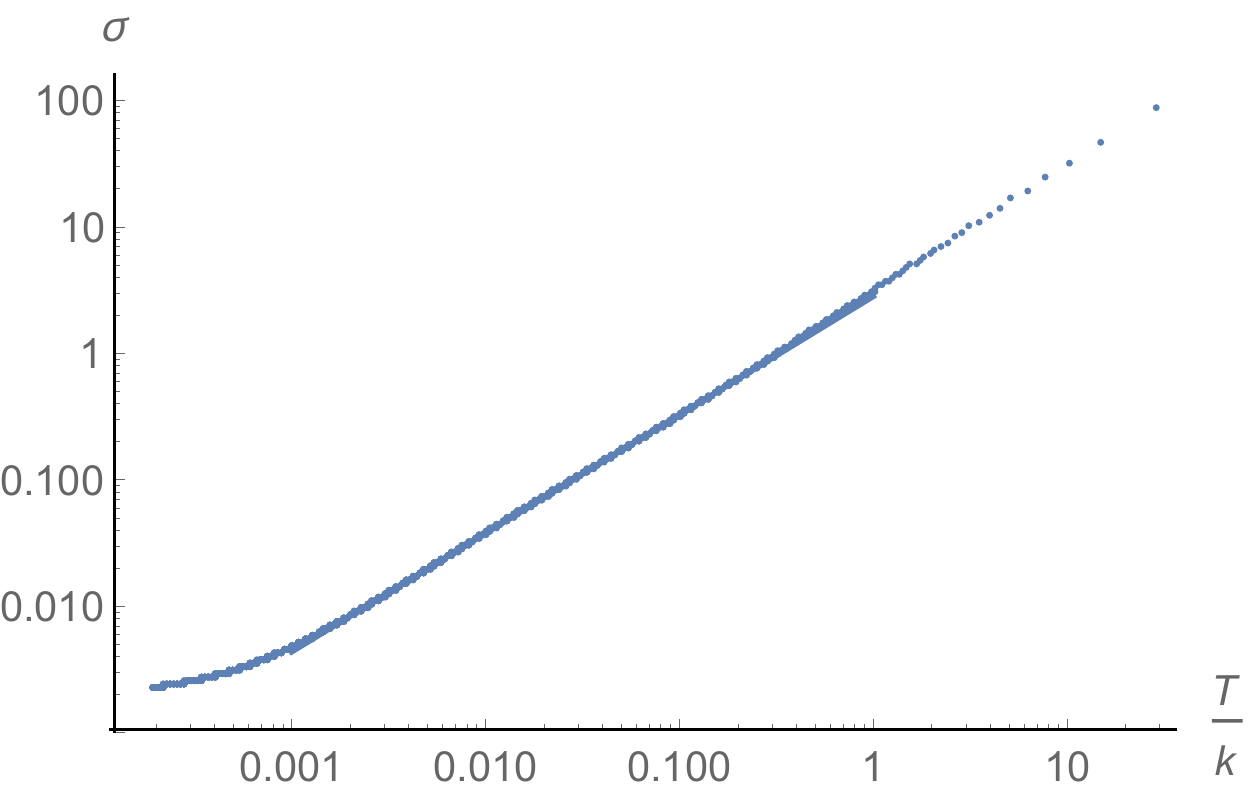}\hfill\includegraphics[scale=0.63]{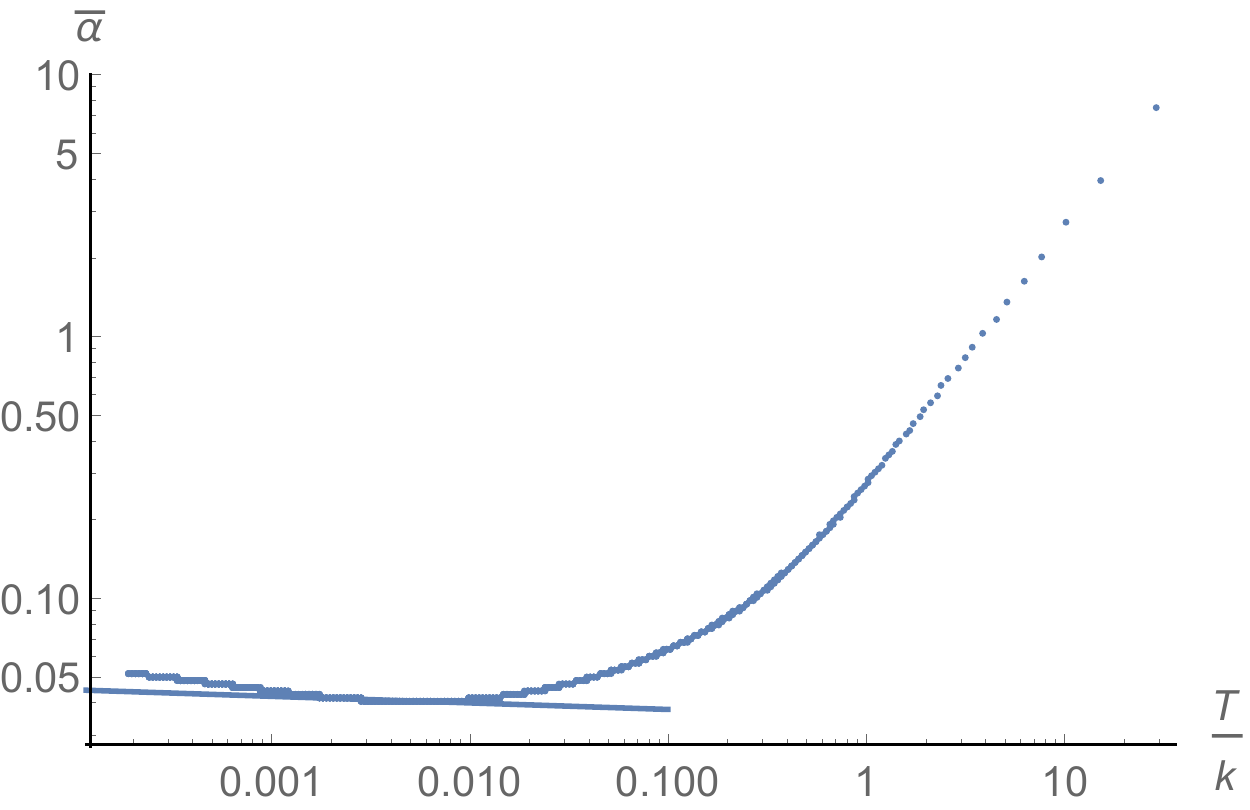}
\caption{ Plot for the conductivities $\sigma $ and $\bar\alpha$ as functions of the reduced temperature $T/k$ for a family of solutions with fixed $\mu/k=0.01$. The straight lines correspond to fits in the intermediate scaling regime. \label{fig3sl2r}}
\end{center}  
\end{figure}

In Figure \ref{fig3sl2r} we plot these coefficients for the family of solutions presented in Figure  \ref{fig2sl2r}. From the plot we see that at large temperatures the coefficients scale as $\sigma/k\sim \bar \alpha\sim T/k$. This is again the expected result for $CFT_4$. As we lower the temperature we find that in the intermediate scaling regime the conductivities also follow power laws. For $\sigma/k $ we observe that still goes linear with the reduced temperature while $\bar\alpha$ remains constant. In the low temperature regime we find  $\sigma/k\sim (T/k)^{1/4}$ and $\bar\alpha\sim 0.052$. 

Now lets continue with the computation of the remaining conductivities. In order to do that  we perturb the Einstein-Maxwell solutions shown in Figure \ref{fig2sl2r} by  
\bea
\delta A&=& (-t \delta f_1(r)+\delta a_x(r))dx\,,\nn\\
\delta ds^2&=& (t \delta f(r)+h_{tx}(r))dxdt + h_{rx}dxdr\,.
\eea
Following as in the previous sections we obtain expressions for $J$ and $Q$ that does not depend on the radial direction and can be just expressed as function of IR data
\bea
J&=&\frac{f_0^2 k^2 r_h \left(6 \left(2 h_0^6+1\right) E -a_{t_1} \zeta  \left(4 h_0^6+1\right) r_h\right)-2 a_{t_1}
h_0^8 r_h^3 \left(\zeta r_h \left(a_{t_1}^2-12 f_0^2\right)+6 a_{t_1} E \right)}{6 f_0^2 h_0^2 \left(2
 h_0^6+1\right) k^2}\,,\nn\\
Q&=&\frac{r_h \left(2 h_0^8 r_h^2 \left(a_{t_1}^2-12 f_0^2\right)+f_0^2 \left(4 h_0^6+1\right) k^2\right) \left(2 h_0^8
  r_h \left(\zeta  r_h \left(a_{t_1}^2-12 f_0^2\right)+6 a_{t_1} E \right)+\zeta  f_0^2 \left(4 h_0^6+1\right)
   k^2\right)}{72 f_0^2 h_0^{10} \left(2 h_0^6+1\right) k^2}\,,
\eea
where we erased the temporal dependence fixing
\bea
\delta f_1(r)&=&E+\zeta A_t\,,\nn\\
\delta f(r)&=&2r^2\zeta f^2 g\,,
\eea
and we used that one of the Einstein equations is
\be
\delta h_{rx}=-\frac{2  k^2 r h^8 A_t' \left(\zeta  r A_t'+6 \zeta A_t+6 E \right)+\zeta  f^2 \left(12 r^2 h^6 \left(-r^2 g h'^2+r
  g h h'+2 (g-1) h^2\right)+\frac{1}{k^2}\right)}{3 r f^2 g}\,.
\ee

Then, the remaining transport coefficients read
\bea
\alpha &=&\frac{4 \pi  a_{t_1}h_0^6 r_h^3}{k^2 \left(2 f_0 h_0^6+f_0\right)}\,,\nn\\
\bar\kappa &=&\frac{\pi  r_h^2 \left(-2 h_0^8 r_h^2 \left(a_{t_1}^2-12 f_0^2\right)-f_0^2 \left(4 h_0^6+1\right) k^2\right)}{3
  h_0^2 k^2 \left(2 f_0 h_0^6+f_0\right)}\,.
\eea
We recover the expected result $\alpha=\bar\alpha$. The thermal conductivity at zero electric current is written for this gravity solution as
\be 
\kappa=\bar\kappa-\frac{\alpha\bar\alpha T}{\sigma}=\frac{2 \pi  f_0 h_0^8 r_h^4 \left(12 f_0^2-a_{t_1}^2\right)-\pi f_0^3 \left(4 h_0^6+1\right) k^2 r_h^2}{6
 a_{t_1}^2 h_0^{10} r_h^2+3 f_0^2 \left(2 h_0^6+1\right) h_0^2 k^2}\,.
\ee

\begin{figure}[htb]
\begin{center}  
\includegraphics[scale=0.7]{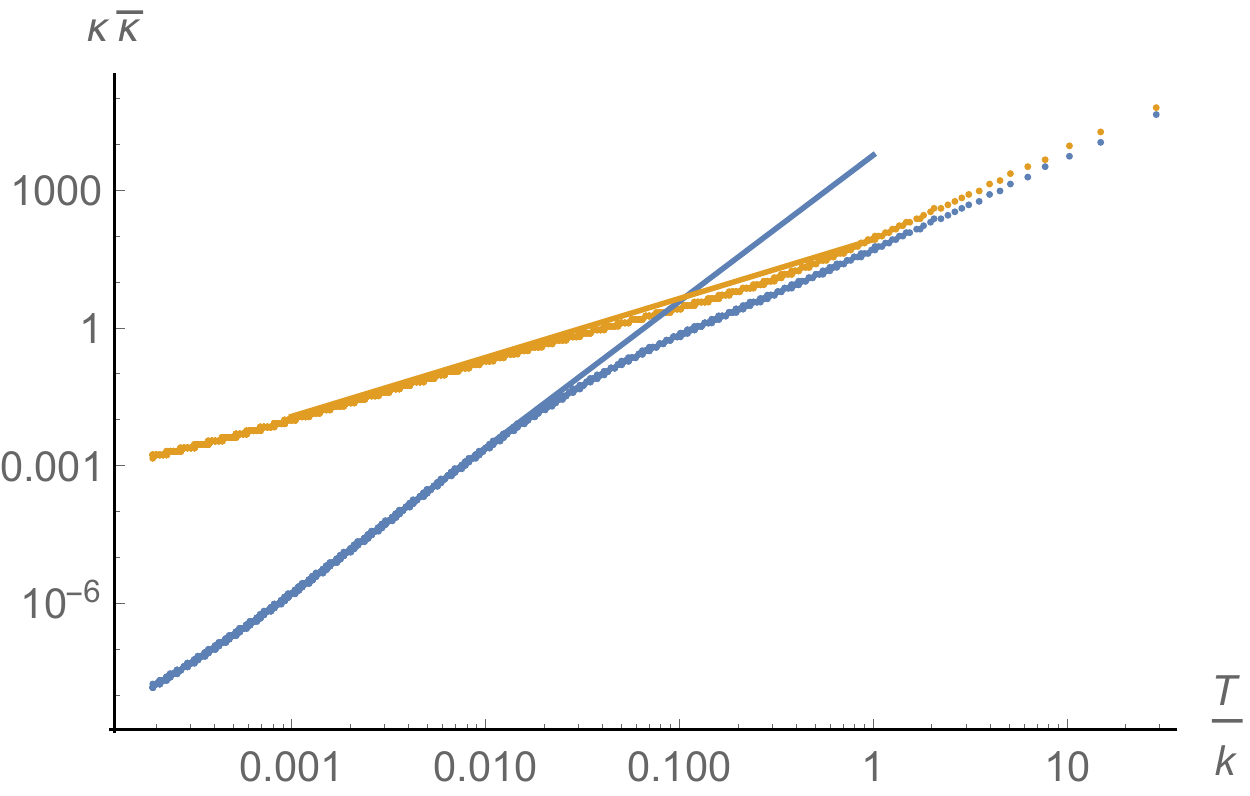}
\caption{ Conductivities $\bar\kappa $ (blue) and $\kappa$ (orange) as functions of the reduced temperature $T/k$ for a family of solutions with fixed $\mu/k=0.01$. \label{fig4sl2r}}
\end{center}  
\end{figure}

In Figure \ref{fig4sl2r} we compare these two quantities and their behaviour as functions of the reduced temperature. Again, we recover the expected $\kappa\sim\bar\kappa\sim (T/k)^2$ at high temperatures. As we lower the temperature and  enter into the intermediate regime, we see that $\kappa$ and $\bar\kappa$ start to scale differently, $\kappa\sim (T/k)^{3.19}$ and $\kappa\sim (T/k)^{1.28}$. In the low temperature regime we find $\kappa\sim (T/k)^{2.53}$ and $\kappa\sim (T/k)^{0.97}$.

\newpage

\section{Summary and Future directions}
\label{sec5}

We constructed a class of black holes that implements an RG flow which pass trough intermediate scaling solutions within Einstein-Maxwell-AdS theory. These intermediate scalings are anisotropic and are triggered by the fact that the theory lives in a curved geometry associated to a Bianchi geometry (\ref{bianchist}). An interesting characteristic of our solutions is that in the intermediate regime different spatial directions scale differently giving a generalized idea o Lifshitz scaling. We sumarize the scalilngs of the space time for charged solutions in the following table

\begin{table}[h]
\centering 
\begin{tabular}{@{}|l|l|l|l|l|l|l|l|l|l|l|l|l|@{}}
\toprule
\multirow{2}{*}{} & \multicolumn{3}{c|}{$t$} & \multicolumn{3}{c|}{$\omega_1$} & \multicolumn{3}{c|}{$\omega_2$} & \multicolumn{3}{c|}{$\omega_3$} \\ \cmidrule(l){2-13} 
                         & UV    & Inter.    & IR   & UV    & Inter.    & IR   & UV    & Inter.    & IR   & UV      & Inter.     & IR     \\ \midrule
Solv                  &   1    &   1    &    1  &     1  &          0  &  0      &  1    &        1       &  0  &   1    &         0    &     0   \\ \midrule
Nil                     &    1   &   11/8 &1   &    1   &       1    &       0   &  1     &       1         &   0  &  1   &        3/2  &    0   \\ \midrule
$SL_2({\cal R})$ &   1  &   1     &   1  &    1   &      0     &    0  &    1   &         0         &   0  &    1    &     1  &       0     \\ \bottomrule
\end{tabular}
\end{table}

In the lattice induced scalings presented in  \cite{Horowitz:2012gs,Donos:2017ljs,Donos:2017sba}, all spatial directions have the same scalings. We believe that this might be a consequence of  considering square lattices and it would be nice to check if different lattice shapes will induce anisotropic scalings like the ones we present here.

The DC conductivities also reflect the fact that there exist new scaling solutions at intermediate temperatures. In the following table we summarize the powers of $T/k$ obtained for the different conductivities and black hole solutions

\begin{table}[h]
\centering 
\begin{tabular}{@{}|l|l|l|l|l|l|l|l|l|l|l|l|l|@{}}
\toprule
\multirow{2}{*}{} & \multicolumn{3}{c|}{$\sigma/k$} & \multicolumn{3}{c|}{$\alpha$} & \multicolumn{3}{c|}{$\kappa$} & \multicolumn{3}{c|}{$\bar\kappa$} \\ \cmidrule(l){2-13} 
                  & UV    & Inter.    & IR   & UV    & Inter.    & IR   & UV    & Inter.    & IR   & UV      & Inter.     & IR     \\ \midrule
Solv              &   1    &         1.88         &    0.14  &     1  &          0.39        &  0    &  2     &          1.66        &   1.5   &     2    &                 1.66  &     1   \\ \midrule
Nil               &    1   &        1.38          &0   &    1   &          -0.44        &  0   &  2     &         0.71         &    1.45  &      2   &          0.63       &    0.65    \\ \midrule
$SL_2({\cal R})$  &   1    &            1      &    0.25  &    1   &        0          &    0  &    2   &         3.19         &    2.53  &    2     &     1.28  & 0.97       \\ \bottomrule
\end{tabular}
\end{table}

Let us finish by discussing some posible future directions. A first direction would be to check the generality of our results by studying other possible boundary geometries. In reference \cite{Donos:2014gya}, the authors studied the Einstein equations with a Bianchi $VII_0$ deformation and found boomerang RG flows. It would be interesting to see wether one can find also intermediate scalings in the large helical deformation regime. Another promising geometry to explore would be the squashed sphere Bianchi type $IX$.

Another interesting direction would be to explore the effect of $U(1)$ symmetry breaking in the scalings. In this direction, a holographic superconductor living on an helical background was studied in \cite{Erdmenger:2015qqa}. There the helix is supported by a Proca field. It would be an interesting excercise to repeat their study within the simpler Einstein-Maxwell context.

Finally it would be interesting to see how these scalings appear in the weak coupling limit by studying a theory with simple field content living in one of the spaces studied above. It would be nice to find in some toy model how the anisotropic scalings appear in perturbation theory.

\section*{Acknowledgements}

We would like to thank Gonzalo Torroba for insightful discussions. ISL acknowledges hospitality from IFLP, where part of this work was done.
ISL is a  CONICET fellow.

\end{document}